\documentclass{aa}
\usepackage[varg]{txfonts}
\usepackage{graphicx}				
\usepackage{xcolor}
\PassOptionsToPackage{hyphens}{url}\usepackage{hyperref}
\usepackage{url}

\newcommand{\centralbibliography}{/Users/pino/Documents/bibliography}
\bibpunct{(}{)}{;}{a}{}{,}

\newcommand{\pyeta}{$^\pi \eta$}
\newcommand{\harps}{$\left(4\nu+\delta\right)$}
\newcommand{\espresso}{$\left(4\nu\right)$}
\newcommand{\gianoone}{$\left(2\nu+\delta\right)$}
\newcommand{\gianotwo}{$\left(2\nu\right)$}
\newcommand{\gianothree}{$\left(\nu+\delta\right)$}

\begin{document}
\title{Diagnosing aerosols in extrasolar giant planets with cross-correlation function of water bands}
\author{Lorenzo Pino \inst{1,2}
\and David Ehrenreich \inst{1}
\and Romain Allart \inst{1}
\and Christophe Lovis \inst{1}
\and Matteo Brogi \inst{3}
\and Matej Malik \inst{4}
\and Valerio Nascimbeni \inst{2,5}
\and Francesco Pepe \inst{1}
\and Giampaolo Piotto \inst{2}
}

\institute{ Observatoire astronomique de l’Universit\'e de Gen\`eve, Universit\'e de Gen\`eve, 51 chemin des Maillettes, CH-1290 Versoix, Switzerland
\and Dipartimento di Fisica e Astronomia `Galileo Galilei', Univ. di Padova, Vicolo dell’Osservatorio 3, Padova I-35122, Italy
\and Department of Physics, University of Warwick, Coventry CV4 7AL, UK
\and University of Bern, Center for Space and Habitability, Sidlerstrasse 5, CH-3012, Bern, Switzerland
\and INAF - Osservatorio Astronomico di Padova, vicolo dell'Osservatorio 5, I-35122, Padova, Italy}

						
\abstract{
Transmission spectroscopy with ground-based, high-resolution instruments provides key insight into the composition of exoplanetary atmospheres. Molecules such as water and carbon monoxide have been unambiguously identified in hot gas giants through cross-correlation techniques. A maximum in the cross-correlation function (CCF) is found when the molecular absorption lines in a binary mask or model template match those contained in the planet. Here, we demonstrate how the CCF method can be used to diagnose broad-band spectroscopic features such as scattering by aerosols in high-resolution transit spectra. The idea consists in exploiting the presence of multiple water bands from the optical to the near-infrared. We produced a set of models of a typical hot Jupiter spanning various conditions of temperature and aerosol coverage. We demonstrate that comparing the CCFs of individual water bands for the models constrains the presence and the properties of the aerosol layers. The contrast difference between the CCFs of two bands can reach $\sim100~\mathrm{ppm}$, which could be readily detectable with current or upcoming high-resolution stabilized spectrographs spanning a wide spectral range, such as ESPRESSO, CARMENES, HARPS-N+GIANO, HARPS+NIRPS, SPIRou or CRIRES+.
}
\maketitle	
\section{Introduction}
\label{sec: intro}
High-resolution ($R\sim10^{5}$), ground-based transmission spectroscopy offers a unique insight into exoplanetary atmospheres. At such high resolution	 (1) telluric contamination, which dramatically hampers lower resolution, ground-based observations \citep{Sing2015, Gibson2017}, is better removed \citep{Astudillo-Defru2013, Allart2017}; (2) the resolved cores of the alkali atoms sound up to the base of the thermosphere ($P\lesssim10^{-9}~\mathrm{bar}$; \citealt{Wyttenbach2015, Wyttenbach2017}); (3) individual absorption lines of molecular bands are spectrally resolved and chemical species, such as $\mathrm{H_2O}$ and $\mathrm{CO}$, are uniquely identified \citep{Snellen2010, DeKok2014, Brogi2018}.\\
Single molecular bands have been detected with high-resolution transmission spectroscopy in the NIR by exploiting cross-correlation techniques (e.g. \citealt{Brogi2016}). \cite{Hoeijmakers2015}, \cite{Allart2017} and \cite{Esteves2017} have proposed an extension of this method to the optical bands. However, optical to NIR high-resolution spectrographs are only available from the ground\footnote{The optical and NIR spectral range will be covered by JWST NIRSpec, which will provide a resolving power of $R\sim10^3$. This is by two orders of magnitude lower than the resolving power that ground-based facilities can provide.}, thus the effectiveness of these techniques has been limited by the presence of Earth atmosphere.\\
One aspect of the problem is that high-resolution transmission spectroscopy requires a normalization of the flux of the spectra acquired during the night, that can vary due to airmass or instrumental effects \citep{Snellen2008, Wyttenbach2015, Heng2015, Wyttenbach2017}. During the process, the absolute level of the absorption of the planet is lost, and broadband features such as those due to aerosols could be removed from the spectra. Indeed, the presence of aerosols on exoplanets was inferred with space-borne HST observations (e.g. \citealt{Charbonneau2002, Ehrenreich2014, Kreidberg2014Nat, Wakeford2017W101}, but see also \citealt{Snellen2004, DiGloria2015}) or lower resolution spectrophotometric observations (e.g. \citealt{Bean2010, Nascimbeni2013, Lendl2016}), revealing that they are very common on hot Jupiters \citep{Sing2016, Barstow2017} and smaller Saturn- and Neptune-size planets. \cite{Brogi2017}, who did not treat the effect of aerosols explicitly, and \cite{Pino2018} demonstrated that a combined analysis of ground-based, high-resolution and space-borne, low- to medium-resolution transmission spectra can better constrain the chemical and physical conditions of exoplanetary atmospheres.\\
Single spectral features such as the sodium doublet can be used to infer the presence of aerosols \citep{Charbonneau2002, Snellen2008, Heng2016, Wyttenbach2017}. Here we demonstrate quantitatively that high-resolution transmission spectroscopy of molecular bands can be used to distinguish between aerosol-free and aerosol-rich scenarios (as qualitatively noted by \citealt{DeKok2014}). The technique exploits the fact that the opacities of water vapour and aerosols behave differently with wavelength: the former increases with wavelength, while the latter is usually constant or decreasing in the optical and NIR regions of the spectrum. The relative contrast of water bands can thus be used as a diagnostic of the presence of aerosols. At high resolution, the contrast of water bands can be measured through their cross-correlation functions (CCFs). This is obtained by cross-correlating each water band with a binary mask, and can thus be regarded as the average line profile within the band.\\
We use the \pyeta{} code (\citealt{Pino2018}; see also \citealt{Ehrenreich2006, Ehrenreich2012, Ehrenreich2014}) to build a grid of models representative of a hot Jupiter in different temperature conditions, containing aerosols at different altitudes; build the CCFs of the grid with a set of binary masks representing water bands of interest (following the technique introduced by \citealt{Allart2017}); discuss how the relative contrast of different water absorption bands is an indicator of the cloudiness of an atmosphere.\\
Our technique is suitable for applications with the roster of current and next generation high-resolution ($R\gtrsim 50\,000$), stabilized, optical-to-NIR Echelle spectrographs with extended spectral coverage ($\Delta\lambda\gtrsim 2\,000~\mathrm{\AA}$). A non-comprehensive list includes: HARPS + NIRPS, GIARPS (GIANO + HARPS-N), ESPRESSO, CARMENES, SPIRou, ARIES, IGRINS, IRD, HPF, iLocater, PEPSI, EXPRES, NEID \citep{Mayor2003, Conod2016, Oliva2006, Cosentino2012, Claudi2017, Pepe2010, Quirrenbach2014, Thibault2012, McCarthy1998, Yuk2010, Kotani2014, Mahadevan2012, Crepp2014, Strassmeier2015, Jurgenson2016, Schwab2016}. Of particular interest is the HIRES optical-to-NIR spectrograph for the E-ELT \citep{Zerbi2014, Marconi2016}, HROS \citep{TMT2013} for the TMT and G-CLEF \citep{Szentgyorgyi2012, Szentgyorgyi2014} for the GMT. At longer wavelengths ($\lambda \gtrsim 2~\mu\mathrm{m}$), an extension  of our technique is possible and warranted in view of future instruments such as CRIRES+ \citep{Follert2014}, METIS \citep{Brandl2014}, NIRES-R and GMTNIRS. See \cite{Crossfield2014} and \cite{Wright2017} for a description of most of these instruments.
\section{Methods}
\label{Sec:Methods}
\begin{figure*}
\resizebox{\hsize}{!}{\includegraphics{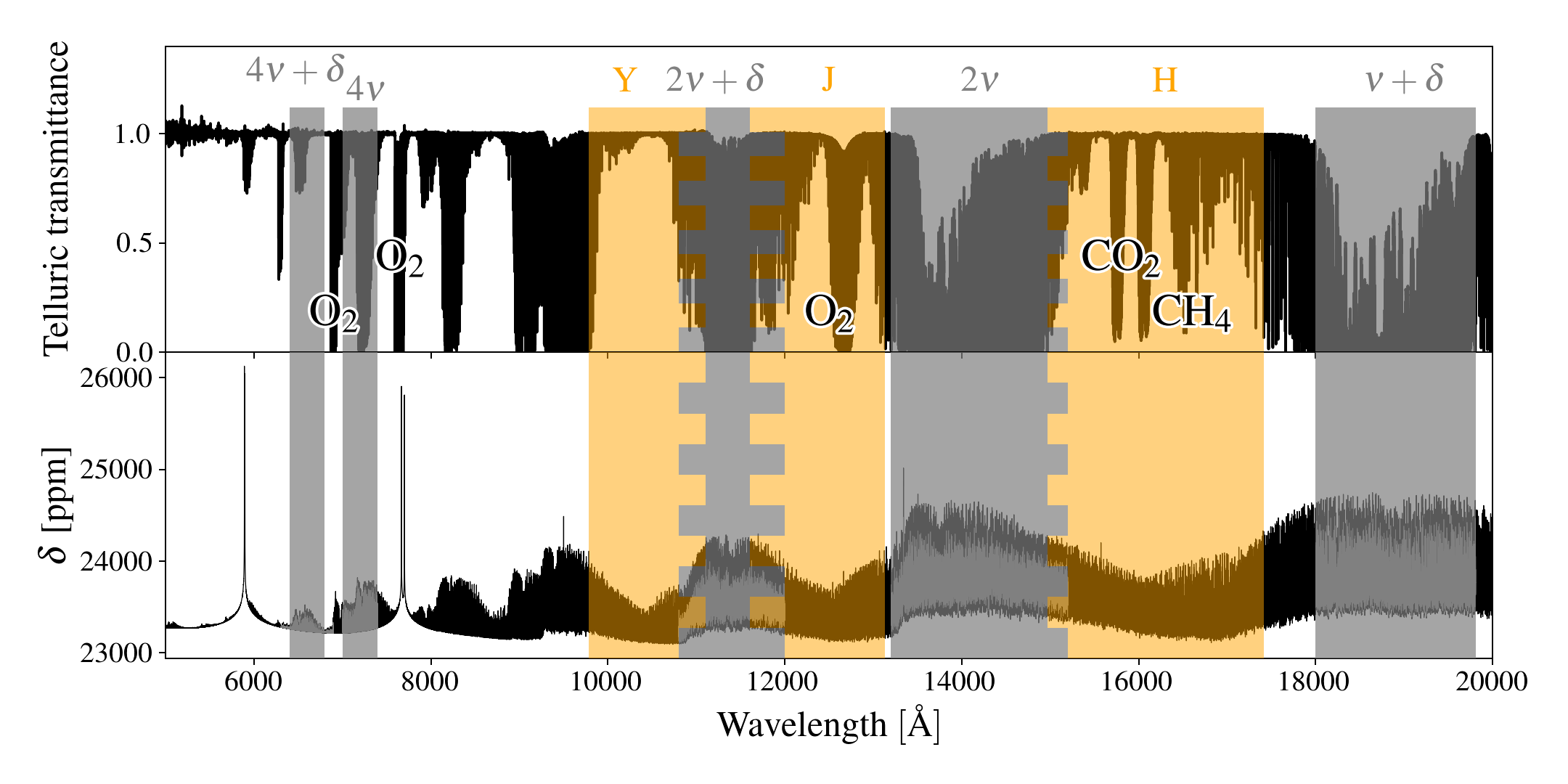}}
\caption{Illustration of the bands used to compute the CCF of water, on-top of the telluric transmittance (upper panel) and an exoplanet transmission spectrum model (lower panel). Gray bands corresponds to water vibrational band. Orange bands correspond to the windows of transparency of Earth atmosphere. The bands partially overlap, as indicated by toothed areas. \textit{Upper panel:} The solid black line represents the transmittance of Earth atmosphere from \citep{Hinkle2003}, ranging from 1 (all incident light transmitted) to 0 (all incident light absorbed). Besides water bands, corresponding to the bands in the bottom panel, molecular oxygen, carbon dioxide and methane show the most prominent features. \textit{Lower panel:} The solid black line represents a theoretical cloud-free isothermal model of HD189733b. Within each water vibrational band, the region of the spectrum in gray represents the spectrum seen through a selected instrument, thus convolved at the appropriate resolving power (see Table \ref{tab:CCF_parameters}). \label{Fig:CCF_parameters}}
\end{figure*}

First, we select the relevant spectral bands. We consider two cases: without telluric contamination (ideal case) and with telluric contamination (realistic case). This yields two sets of bands listed in Tables \ref{tab:CCF_parameters} and \ref{tab:CCF_parameters_tellurics} respectively\footnote{A discussion of the theoretical water spectrum can be found at \url{http://www1.lsbu.ac.uk/water/water_vibrational_spectrum.html} and references therein.} (see also Fig. \ref{Fig:CCF_parameters}). The ideal case informs on the potential of a high-resolution optical to NIR spectrograph mounted on a space-borne or stratospheric facility. While such an instrument will not be delivered within the next decade, similar concepts have already been developed (LUVOIR, \citealt{France2017}; HIRMES\footnote{\url{https://science.nasa.gov/technology/technology-}\\ \url{stories/HIRMES-High-resolution-Mid-infrared-} \\ \url{Spectrometer-SOFIA}}, on-board SOFIA). On the other hand, from the ground, \cite{Allart2017} demonstrated how telluric lines can be efficiently corrected for in the optical. In the NIR, the contamination is more severe. Although telluric correction or self-calibration using time resolution and the measured flux variations \citep{DeKok2013, Birkby2013} both work effectively for unsaturated telluric lines, portions of the spectrum completely saturated, e.g. at the centre of strong water lines, cannot be recovered. We thus select regions where the features in the telluric transmittance spectrum, taken from \cite{Hinkle2003}, remove less than 80\% of the flux, and correction is possible. These bands roughly correspond to the windows of transparency of Earth atmosphere, and we name them accordingly (see Table \ref{tab:CCF_parameters}, and orange regions in Fig. \ref{Fig:CCF_parameters}).
\begin{table*}
\caption{Definition of the water absorption bands. They are named after the vibrational mode that determines their central frequency. Rotational fine structure broadens the absorption bands around this frequency. The vibrational number $\nu$ indicates stretching (a combination of symmetric and asymmetric; both have similar frequencies, thus all their combinations overlap when rotational fine structure is included) and $\delta$ indicates bending. With this naming, the strong water band observed with WFC3 G141 corresponds to the \gianotwo{}. We indicate the reference instrument for which we simulated a CCF for our analysis. In the last two columns, we indicate the resolving power and the resolution element of our simulated observations of each band. The simulated instruments are: HARPS for the \harps{} band; ESPRESSO for the \espresso{}  band; for the NIR bands, we study the performance of an instrument of similar characteristics to GIANO.\label{tab:CCF_parameters}}
\centering
\begin{tabular}{lccccc}
\hline 
Band name & Wavelength range $[\mathrm{\AA}]$  & Instrument simulated & Adopted resolving power & Resolution element\\ 
\hline 
\\
\harps{} & $6\,400$--$6\,800~\mathrm{\AA}$ & HARPS(-N) & $115\,000$~$\left(2.6~\mathrm{km~s^{-1}}\right)$ & $0.01~\mathrm{\AA}$\\ 
\rule{0mm}{0.4cm}
\espresso{} & $7\,000$--$7\,400~\mathrm{\AA}$ & ESPRESSO & $134\,000$~$\left(2.2~\mathrm{km~s^{-1}}\right)$ & $0.01~\mathrm{\AA}$\\ 
\rule{0mm}{0.4cm}
\gianoone{} & $10\,800$--$12\,000~\mathrm{\AA}$ & GIANO-like & $50\,000$~$\left(6~\mathrm{km~s^{-1}}\right)$ & $\lambda/50\,000$\\ 
\rule{0mm}{0.4cm}
\gianotwo{} & $13\,200$--$15\,200~\mathrm{\AA}$ & GIANO-like & $50\,000$ &  $\lambda/50\,000$\\ 
\rule{0mm}{0.4cm}
\gianothree{} & $18\,000$--$19\,800~\mathrm{\AA}$ & GIANO-like & $50\,000$ &  $\lambda/50\,000$\\  
\\
\hline 
\end{tabular}
\end{table*}
\begin{table*}
\caption{Definition of the water absorption bands in the presence of telluric absorption. When we consider telluric contamination, the strongest NIR bands are not observable from the ground. We thus replace those by the Earth transparency windows. We indicate the reference instrument for which we simulated a CCF for our analysis. In the last two columns, we indicate the resolving power and the resolution element of our simulated observations of each band. The simulated instruments are: HARPS for the \harps{} band; ESPRESSO for the \espresso{}  band; GIANO for the NIR bands. \label{tab:CCF_parameters_tellurics}}
\centering
\begin{tabular}{lccccc}
\hline 
Band name & Wavelength range $[\mathrm{\AA}]$  & Instrument simulated & Adopted resolving power & Resolution element\\ 
\hline 
\\
\harps{} & $6\,400$--$6\,800~\mathrm{\AA}$ &HARPS(-N) & $115\,000$~$\left(2.6~\mathrm{km~s^{-1}}\right)$ & $0.01~\mathrm{\AA}$\\ 
\rule{0mm}{0.4cm}
\espresso{} & $7\,000$--$7\,400~\mathrm{\AA}$ & ESPRESSO & $134\,000$~$\left(2.2~\mathrm{km~s^{-1}}\right)$ & $0.01~\mathrm{\AA}$\\ 
\rule{0mm}{0.4cm}
Y band & $9\,800$--$11\,100~\mathrm{\AA}$ & GIANO & $50\,000$~$\left(6~\mathrm{km~s^{-1}}\right)$ &  $\lambda/50\,000$\\
\rule{0mm}{0.4cm}
J band & $11\,600$--$13\,100~\mathrm{\AA}$ & GIANO  & $50\,000$ &  $\lambda/50\,000$\\
\rule{0mm}{0.4cm}
H band & $15\,000$--$17\,400~\mathrm{\AA}$ & GIANO & $50\,000$ &  $\lambda/50\,000$\\ 
\\
\hline 
\end{tabular}
\end{table*}
\noindent We compute the expected CCF contrast in each band, following three steps:
\begin{itemize}
\item Simulation of a theoretical template transmission spectrum of an exoplanet at $R\sim10^6$;
\item Simulation of observations of each water band with a relevant spectrograph (see Table \ref{tab:CCF_parameters}), by convolving at the appropriate resolving power and binning to the instrumental resolution;
\item Computation of the CCF with a binary mask.
\end{itemize}
In the following, we describe these three steps (see \citealt{Pino2018} for a more detailed description of the routines used in the first two steps).
\subsection{High resolution template}
\label{sec:template}
\begin{figure*}
\includegraphics[width=17cm]{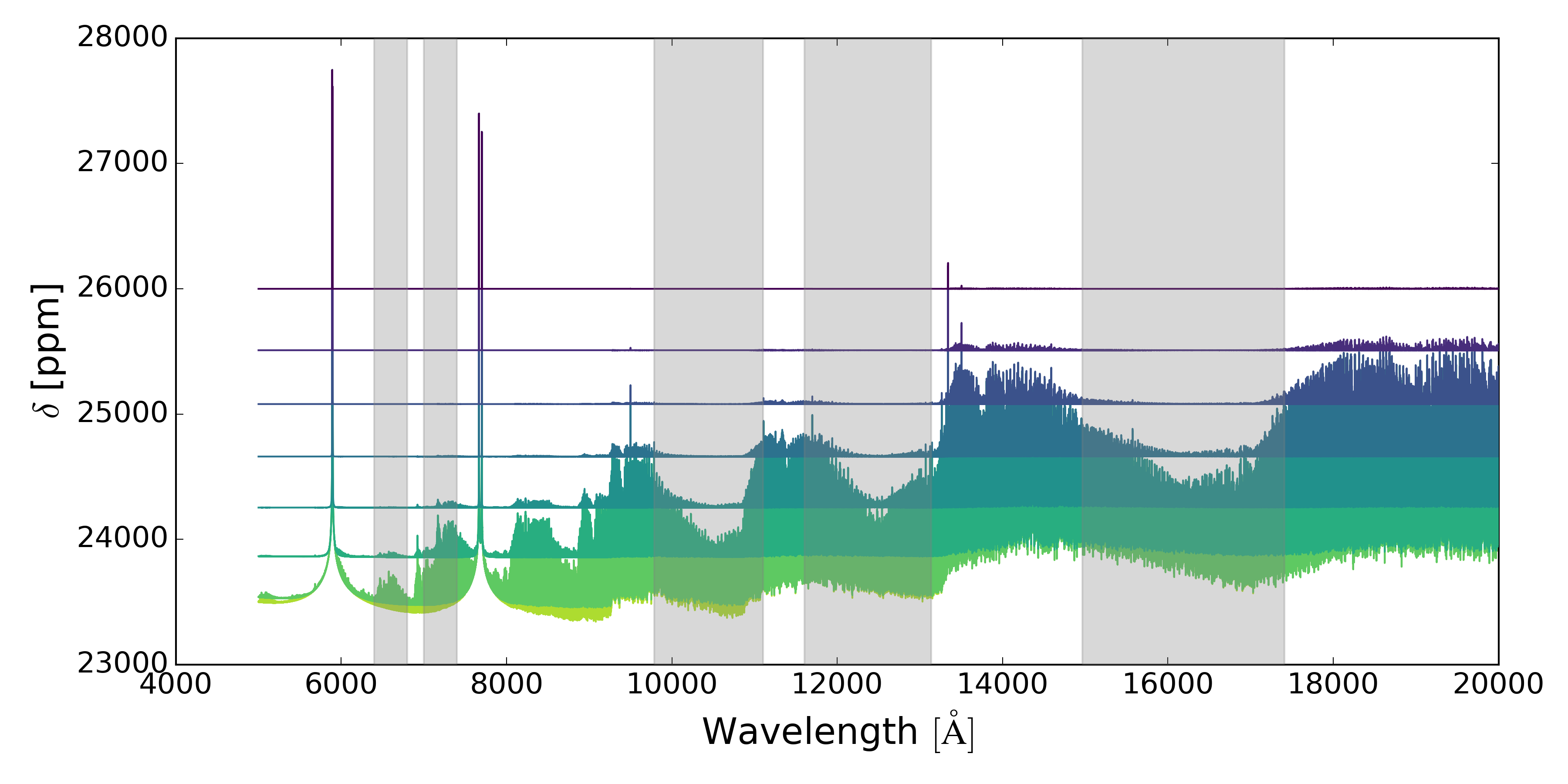}\\
\includegraphics[width=17cm]{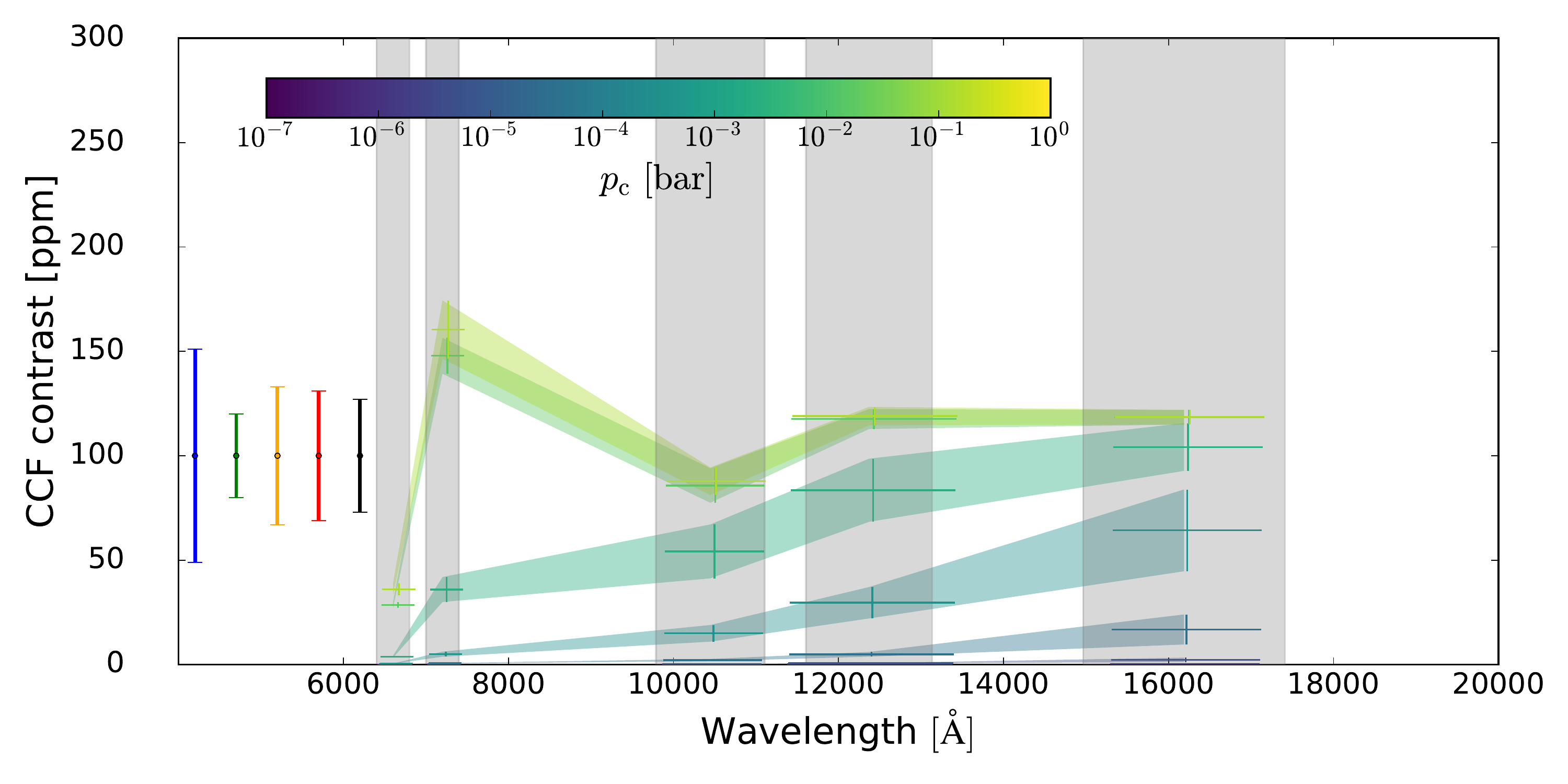}
\caption{CCF contrast as a function of wavelength in the case where telluric contamination is considered, in the presence of a gray opacity source simulating large condensates at various pressure levels. \textit{Upper panel:} High resolution models of a hot Jupiter with $T=1\,700~\mathrm{K}$. Each color corresponds to a different value of $p_\mathrm{c}$, ranging between $10^{-7}~\mathrm{bar}$ and $1~\mathrm{bar}$, and the lightest colors correspond to condensates deeper in the atmosphere. \textit{Lower panel:} CCF contrast as a function of wavelength. The colors are in correspondence with the models shown in the upper panel. For each value of $p_\mathrm{c}$, the temperature range $1\,200$--$2\,300~\mathrm{K}$ is explored. We indicate with a vertical error bar the induced variation in the contrast of the CCF. On the left, the $1\sigma$ precision of HARPS (blue), ESPRESSO (green), GIANO (orange, red and black) for the Y, J and H bands for a single transit are shown. These error bars only account for photon noise, but telluric residuals may introduce additional noise (see Appendix \ref{sec:Detection limits}).\label{Fig:models_clouds_tellurics}}
\end{figure*}

We first use the \pyeta{} 1D line-by-line radiative transfer code \citep{Pino2018} to compute high-resolution model transmission spectra $=\delta(\lambda)=1-f_\mathrm{in}/f_\mathrm{out}$, where $f_\mathrm{in/out}$ are the stellar fluxes registered in- and out-of-transit. A line-by-line approach is necessary to apply the CCF technique, as the molecular lines need to be singularly resolved.
The water cross-section used in the models is calculated from a high-resolution opacity table ($\Delta \overline{\nu}= 0.01~\mathrm{cm}^{-1}$, $R\sim10^6$). The table was obtained with the HELIOS-K routine applied to the HITEMP line-list \citep{Grimm2015}, accurate up to the high temperatures expected in hot Jupiters ($T_{\mathrm{eq}}\gtrsim1\,000~\mathrm{K}$).\\
We assume the planetary parameters of HD189733b, a typical hot Jupiter ($r_\mathrm{p=10\mathrm{bar}} = 1.108~\mathrm{R_J}$, $M_\mathrm{p} = 1.138~\mathrm{M_J}$, $R_\mathrm{\star} = 0.756~\mathrm{R_\odot}$). Besides water (Volume Mixing Ratio, $\mathrm{VMR}=10^{-3}$), we add opacities from sodium ($\mathrm{VMR}=10^{-6}$), potassium ($\mathrm{VMR}=10^{-7}$), and Rayleigh scattering and Collision Induced Absorption (CIA) by $\mathrm{H_2}$. Different parameters are varied to produce the templates:
\begin{enumerate}
\item Aerosol altitude. For illustrating the technique, a simple parametrization of the aerosol deck is sufficient. We set the atmosphere to be opaque at all wavelengths for pressures higher than a threshold $p_\mathrm{c}$. In the spectral range considered ($<2~\mu\mathrm{m}$), this parametrization simulates aerosol particles of size $\gtrsim2~ß\mu\mathrm{m}$. We vary $p_\mathrm{c}$ between $10~\mathrm{bar}$--$10^{-7}~\mathrm{bar}$) in steps of 1 dex (i.e. equally spaced on a logarithmic scale);
\item  Temperature of the atmosphere. We assume isothermal temperature-pressure ($T$-$p$) profiles, at first order adequate to describe water absorption in transmission spectra \citep{Heng2017}. Temperature is varied between $1\,200~\mathrm{K}$ and $2\,300~\mathrm{K}$ in steps of $\sim 200~\mathrm{K}$.
\end{enumerate}
We show a set of templates built for a temperature of $T=1\,700~\mathrm{K}$ in the upper panels of Fig. \ref{Fig:models_clouds} and \ref{Fig:models_clouds_tellurics}. Gray vertical bands correspond to the regions that we analyze in the idealized case and in the realistic case that avoids the peak of telluric contamination, respectively.
\subsection{Convolution with the LSF and binning}
\begin{figure}
\resizebox{\hsize}{!}{\includegraphics{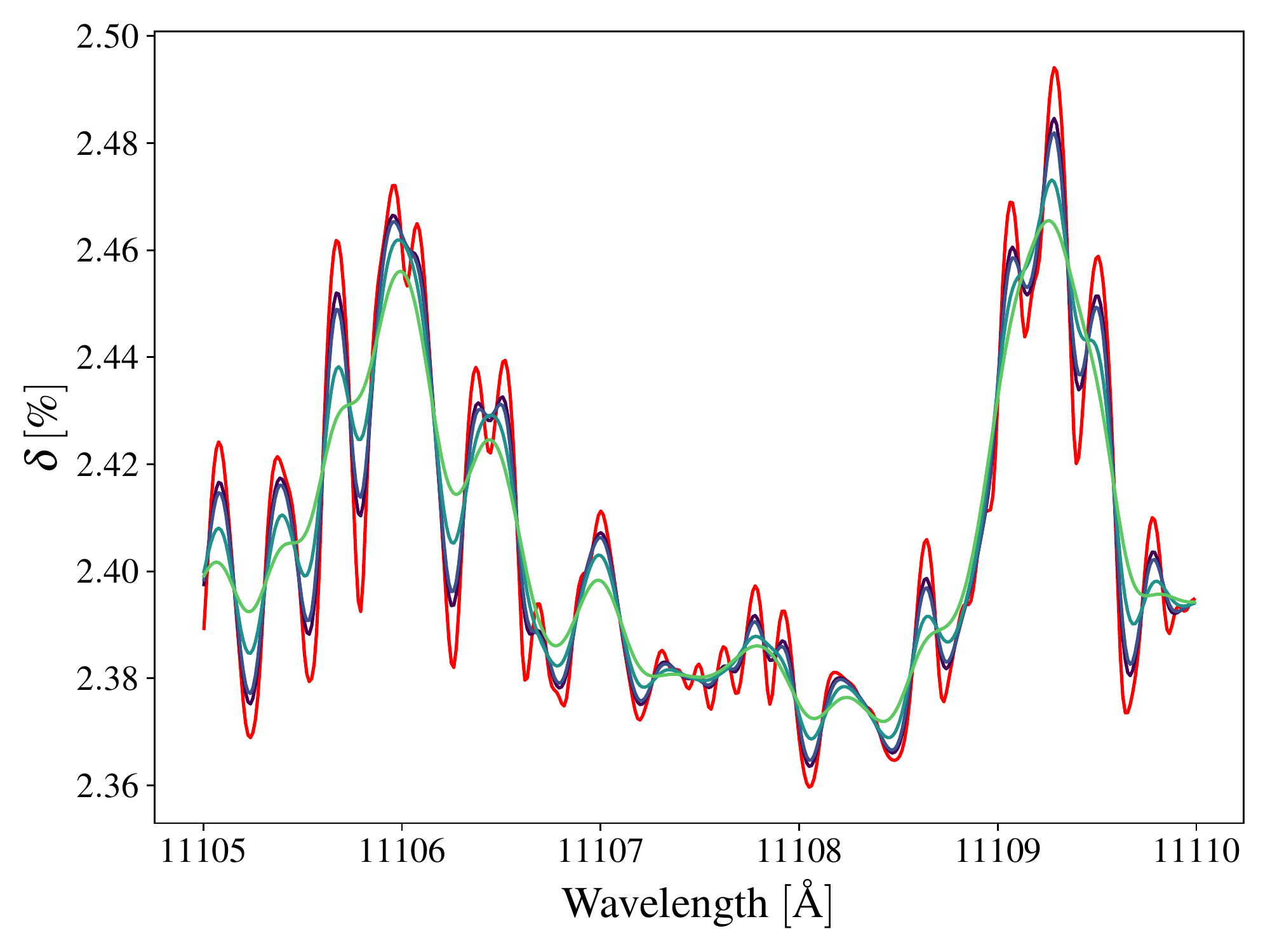}}
\caption{Illustration of the effect of finite resolving power on water lines. The red curve corresponds to a line-by-line model (with a resolution of $R\sim10^6$). From the light green curve to the purple curve we show the same spectrum convolved with Gaussian LSF corresponding to the following resolving powers: at $R=50\,000$ (GIANO), at $R=75\,000$ (similar to CARMENES), at $R=115\,000$ (HARPS/-N) and at $R=134\,000$ (ESPRESSO). At the lowest resolving power the individual molecular lines are not entirely resolved.\label{Fig:resolving_power_effect}}
\end{figure}

When observed with a spectrograph, spectral lines are convolved with the instrumental line spread function (LSF). Even for high-resolution instruments such as HARPS(-N) or ESPRESSO planetary molecular lines are narrower than the instrumental LSF ($\sim2$--$3~\mathrm{km~s^{-1}}$ wide). Indeed, the Doppler width for water lines at typical temperatures of a planetary atmosphere ($500$--$3\,000~\mathrm{K}$) ranges between $0.7~\mathrm{km~s^{-1}}$ and $1.5~\mathrm{km~s^{-1}}$. This convolution effect is strong for resolving powers $<100\,000$, and needs to be taken into account when estimating the contrast of the CCF (see Fig. \ref{Fig:resolving_power_effect}). We convolve lines in each water band with a Gaussian LSF representative of an instrument covering that wavelength range. For illustration we choose to simulate observations with HARPS(-N), ESPRESSO, GIANO (see Tables \ref{tab:CCF_parameters} and \ref{tab:CCF_parameters_tellurics}), but the technique is applicable to any high-resolution spectrograph with enough spectral coverage (see Sec. \ref{sec: intro}).\\
Transmission spectra are built by dividing out stellar spectra taken during the transit and out-of-transit, which are thus individually convolved with the instrumental LSF. \cite{Pino2018} discussed how models should follow the same steps, since mathematically the order of division and LSF convolution cannot be interchanged (see also \citealt{Deming2017}). In this paper, we assume that the stellar host star has no water lines, a reasonable assumption for many cases \citep{Brogi2016}. Thus, the stellar spectrum on the background of planetary water absorption is on average flat, and we can directly convolve the planetary absorption and the instrumental LSF.\\
As discussed in \cite{Pino2018}, a key ingredient of realistic simulations is binning the model to the correct resolution, i.e. the step with which the spectrum is sampled. For the various bands we adopt the resolution elements reported in Tables \ref{tab:CCF_parameters} and \ref{tab:CCF_parameters_tellurics}. For ESPRESSO, we assume that the resolution element is the same as for HARPS(-N).
\subsection{Cross-correlation}
\begin{figure*}
\includegraphics[width=8.5cm]{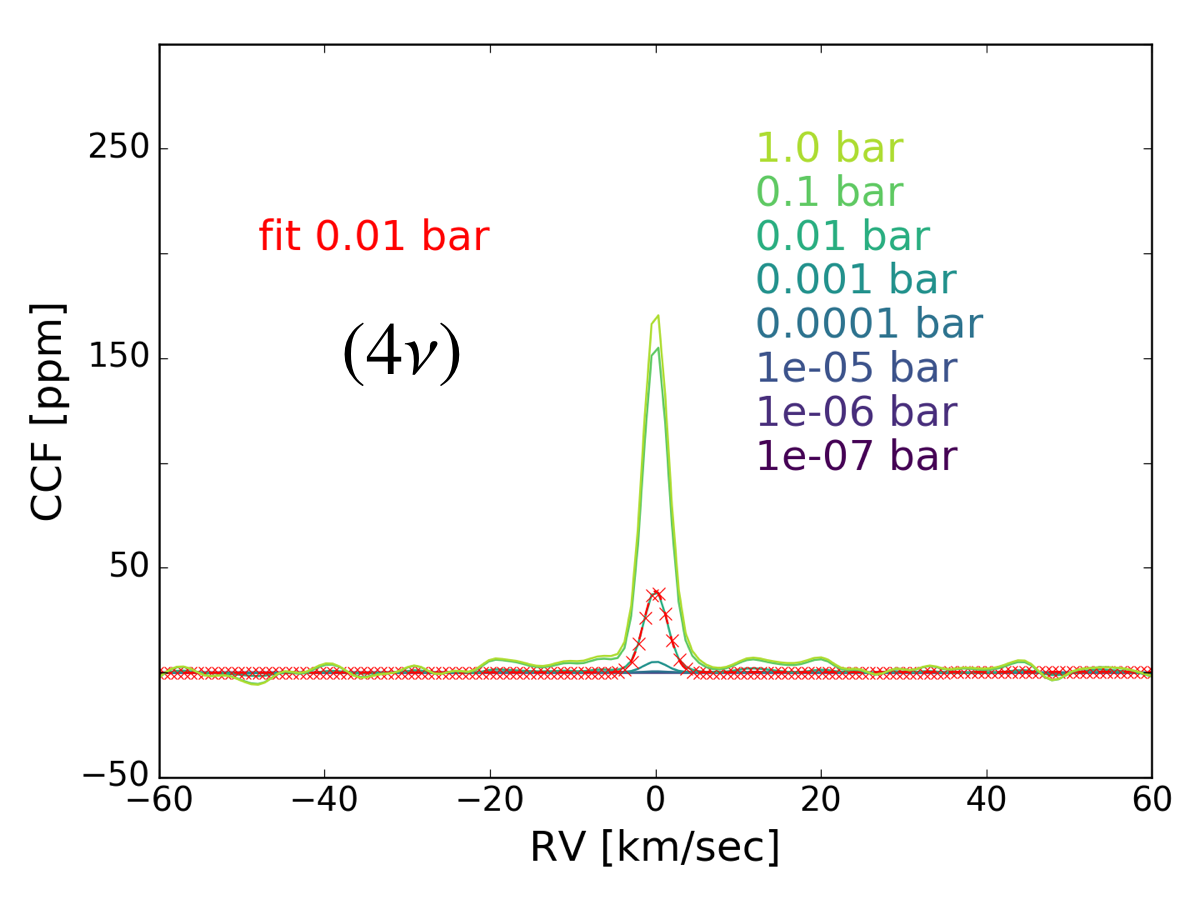}
\includegraphics[width=8.5cm]{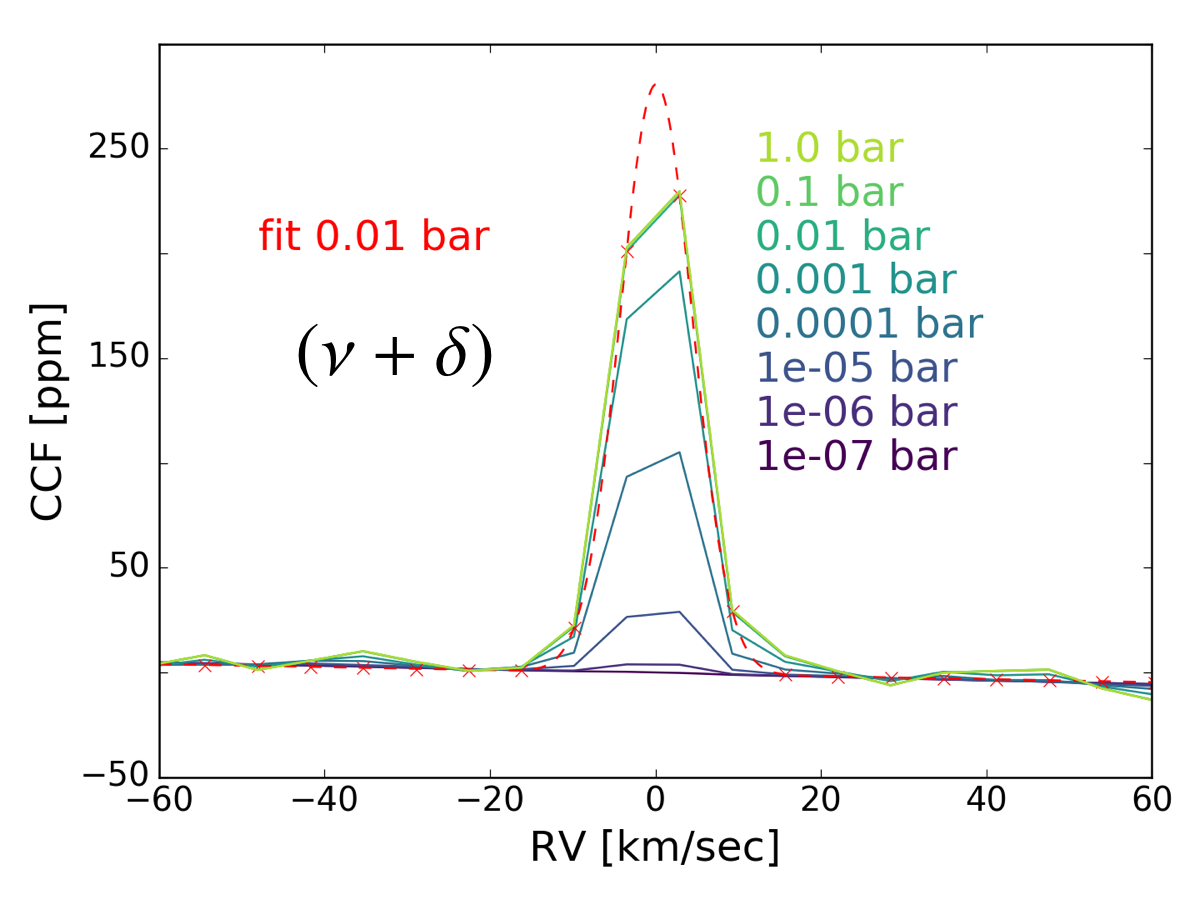}
\caption{CCFs for the \espresso{} and \gianothree{} bands as observed with ESPRESSO and GIANO respectively. We arbitrarily offset the CCFs to place their continuum at 0. Solid lines of different colors from purple to light green show the CCF obtained for a gray aerosol deck located at the pressure indicated in the color matched label. A red dashed line shows the best Gaussian fit to the $p_\mathrm{c}=0.01~\mathrm{mbar}$ case. Red crosses indicate the best fit at the sampling corresponding to the simulation, finer for ESPRESSO, coarser for GIANO.\label{Fig:CCF_examples}}
\end{figure*}

Individual molecular lines are weak and usually buried in the noise. However, we can enhance our detection capabilities by averaging the signal coming from hundreds or thousands of lines. This is achieved by building a Cross-Correlation Function (CCF\footnote{More precisely, we perform cross-correlation as defined in the field of signal processing, which is a measure of similarity between two signals (the binary mask and the data, e.g. \citealt{Press1989}). The terminology in signal processing differs from the terminology in statistics, where the cross-correlation is normalized by the variance of the data. It should be noted that works by other groups, e.g. \cite{Snellen2010}, adopt the CCF as defined in statistics, i.e. normalized by the variance of the two signals. The two definitions of CCF differ just by a normalization factor, and the adoption of one or the other definition should not impact our conclusions.}; \citealt{Baranne1979, Baranne1996, Pepe2002}). To compute the CCF, we follow the method developed by \cite{Allart2017} to search for water in the HARPS optical spectra of HD189733b. The cross-correlation function is defined as:
\begin{equation}
\label{eq:CCF}
CCF_{N,\,T_{ll}}(v)=\frac{1}{N}\sum_{i=1}^N S(\lambda_i)\cdot M\left[\lambda_i \left(1 + \frac{v}{c} \right)\right]~.
\end{equation}
A binary mask $M$ (with an aperture of one pixel, i.e. $0.82~\mathrm{km~s^{-1}}$ in HARPS and ESPRESSO and $6.38~\mathrm{km~s^{-1}}$ in GIANO) is projected on the spectrum $S$. The binary mask corresponds to the theoretical wavelengths $\lambda_i$ of the $N$ lines of the molecule we are looking for, drawn from a line list at a temperature $T_{ll}$. Doppler shifting the mask in the radial velocity ($v$) domain with steps of one pixel, $c$ being the speed og light, allows one to reconstruct the shape of the average molecular line. Following \cite{Allart2017}, we inspect the $-80~\mathrm{km~s^{-1}}/+80~\mathrm{km~s^{-1}}$ range. Since no wavelength shift is introduced in the model, we expect a peak at $0~\mathrm{km~s^{-1}}$.\\
With reference to Eq. (\ref{eq:CCF}), $S$ is one of the water bands defined in Table \ref{tab:CCF_parameters} (without considering telluric absorption) or Table \ref{tab:CCF_parameters_tellurics} (considering it), convolved with the appropriate LSF and binned at the instrumental resolution. We build a dedicated mask for each water band starting from the HITEMP line list. Within the wavelength range of each water absorption band we select the $800$ lines with strongest spectral line intensity at a temperature $T_{ll}=1\,200~\mathrm{K}$\footnote{To rescale the lines intensity we used the partition function provided by HITRAN: \texttt{http://hitran.iao.ru/partfun}. Note also that the spectral line intensity parameter of HITEMP represents the area below the spectral line. If a strong line is also broadened it may be less optimal for the CCF analysis than a weaker but narrow line. We consider this effect negligible when selecting 800 lines.}, typical of the terminator region of hot Jupiters. We note that the choice of the mask is not unique. However, the specific choice we made does not impact the conclusions of the study\footnote{This was verified by repeating the whole analysis with a mask at a temperature of $2\,100~\mathrm{K}$, and by keeping the $400$ and $1\,600$ most intense lines (in place of $800$).}.\\
The CCF of a water absorption band is rich in information, as it encodes the average line profile within the band. In this work, we are only interested in the line contrast. We thus fit a Gaussian profile $f(v)$ to our computed CCF of the form
\begin{equation}
\label{Eq:CCF_profile}
f(v)=C + B\cdot v + \dfrac{A}{\sqrt{2\pi\sigma}}\exp\left(-\dfrac{(v - v_0)^2}{2\sigma^2}\right)\ ,
\end{equation}
leaving $C$, $B$, $A$, $\sigma$ and $v_0$ as free parameters. The contrast of the CCF (i.e. of the average line in the absorption band) is then $f(v_0) - \left(B\cdot v_0 + C\right)$.\\
We show examples of simulated CCFs in Fig. \ref{Fig:CCF_examples}, for the \espresso{} and \gianothree{} bands, observed with ESPRESSO and GIANO respectively. The different resolving power of the observations is clear from the different width of the CCFs. Furthermore, the sampling of the CCF is determined by the sampling of the instrument, which is higher for ESPRESSO (in order to oversample the narrower LSF). We stress that the contrast of the CCF is also dependent on the resolving power (since energy is conserved), and it is thus vital to know well the instrumental profile (hence the choice of stabilized spectrographs with oversampled LSFs). A red dashed line shows the fit to the case of $p_\mathrm{c}=0.01~\mathrm{mbar}$ performed using Eq. \ref{Eq:CCF_profile}. The fit is in good agreement with the simulations. Similar fits are performed for each simulated CCF (not shown).
\begin{figure*}
\includegraphics[width=8.5cm]{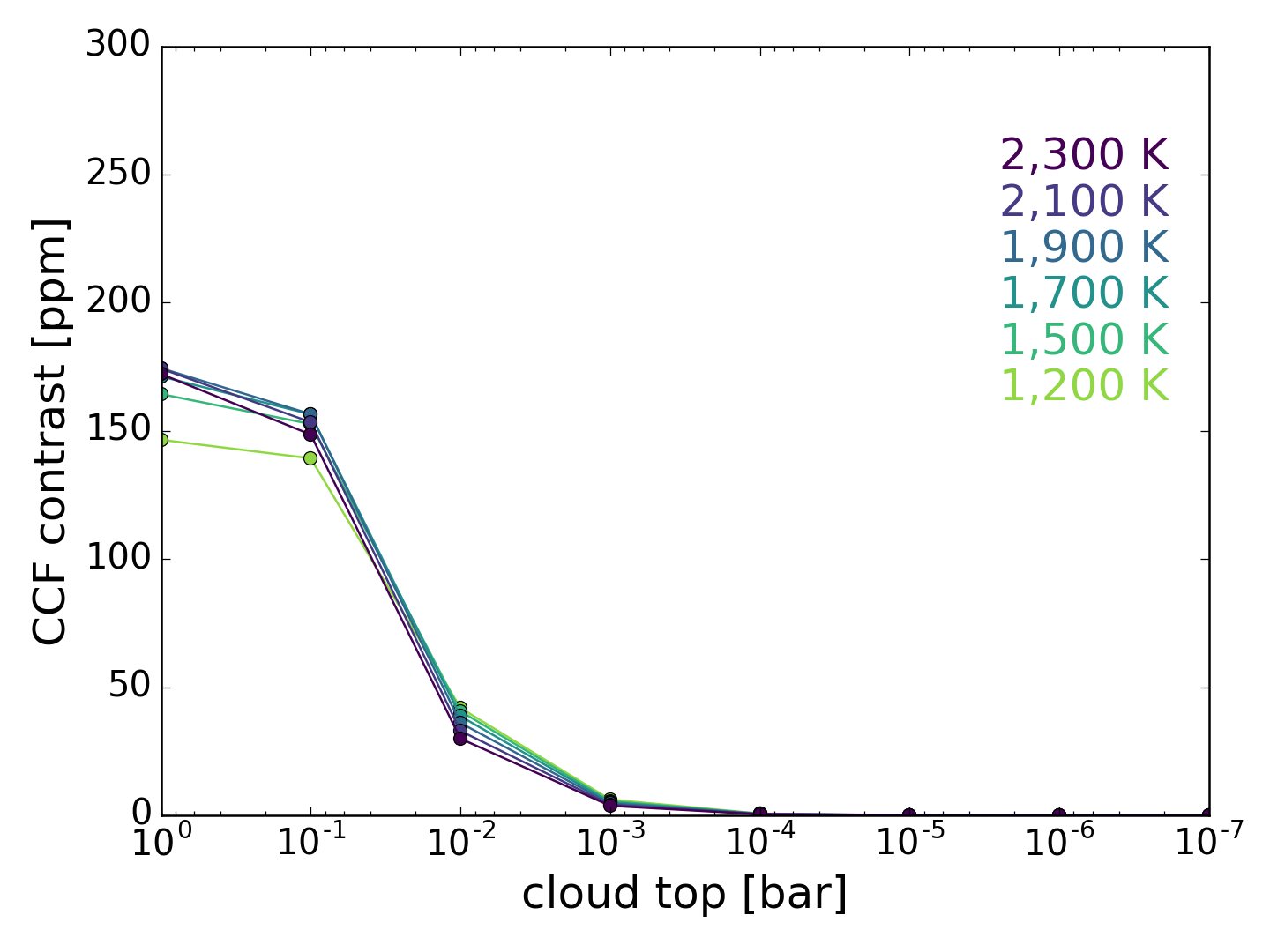}
\includegraphics[width=8.5cm]{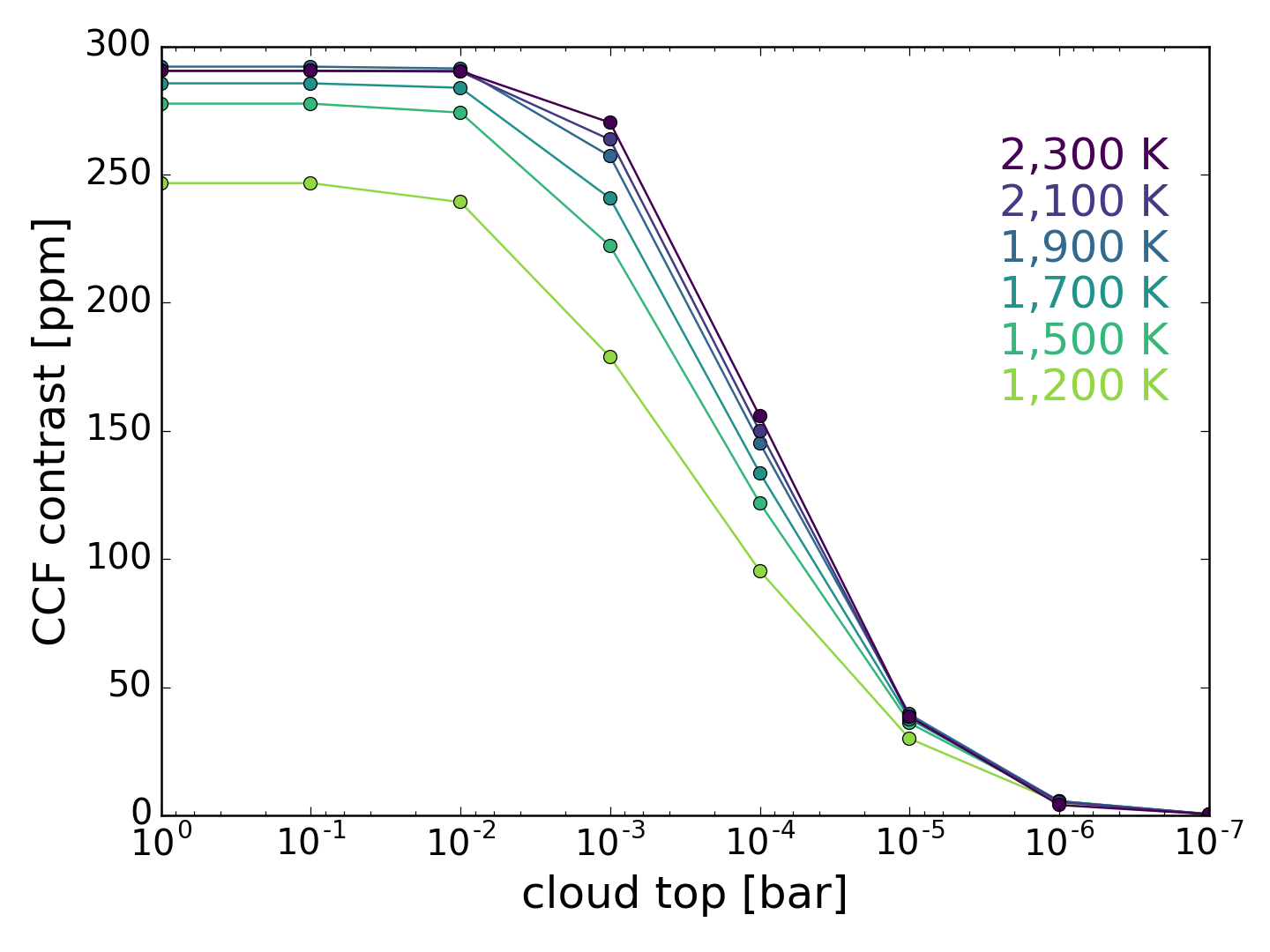}
\caption{\label{Fig:CCF_contrast} Contrast of the CCF as a function of altitude of aerosols (expressed as pressure of the cloud top). Lighter colors indicate colder models, darker colors indicate hotter models (between $1\,200$ and $2\,300~\mathrm{K}$), as indicated by the colour-matched labels. \textit{Left panel:} \espresso{}{} band, as observed with ESPRESSO. A sharp decrease in the contrast of the CCF happens for aerosols higher than $10^{-1}~\mathrm{bar}$ \textit{Right panel:} \gianothree{} band, as observed with GIANO. A sharp decrease in the contrast of the CCF is caused by aerosols higher than $10^{-3}~\mathrm{bar}$, thus this water band is less muted by aerosols compared to the \espresso{}{} band.}
\end{figure*}
\begin{figure*}
\includegraphics[width=17cm]{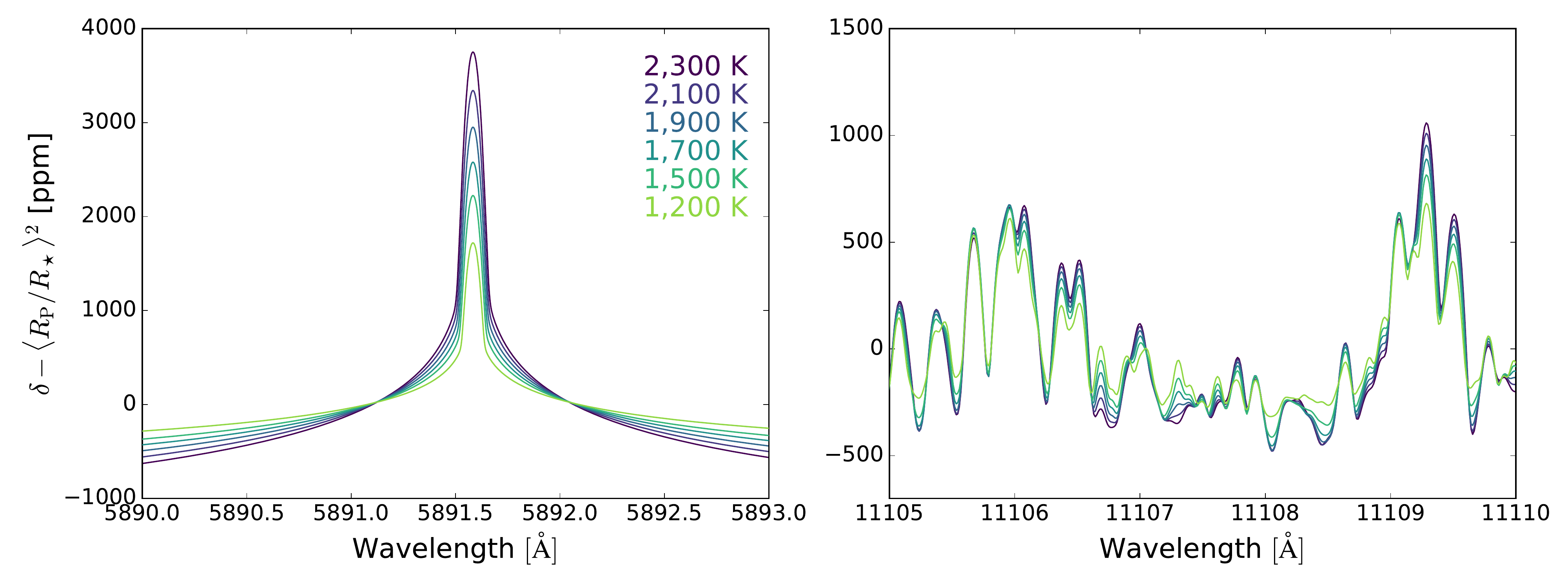}
\caption{Temperature dependence of the high-resolution transmission spectrum of a hot Jupiter. The temperature varies between $1\,200~\mathrm{K}$ and $2\,300~\mathrm{K}$ in steps of $\sim200~\mathrm{K}$ (from light to dark). \textit{Left panel:} Zoom on one line of the sodium doublet. The contrast of the sodium feature in the hottest atmosphere (purple) is approximately twice its contrast in the coldest atmosphere (light green). This behaviour is explained through the linear dependence of atmospheric signal on temperature through the scale height. \textit{Right panel:} Zoom on a region of the \gianoone{} water band. Even if in general the hottest transmission spectrum has a higher contrast compared to the coldest one, the dependence on temperature is non-linear. Some specific lines are actually stronger in a colder atmosphere (e.g. close to $11\,107.5~\mathrm{\AA}$).\label{Fig:temperature_dependence_on_models}}
\end{figure*}	
\begin{figure*}
\includegraphics[width=17cm]{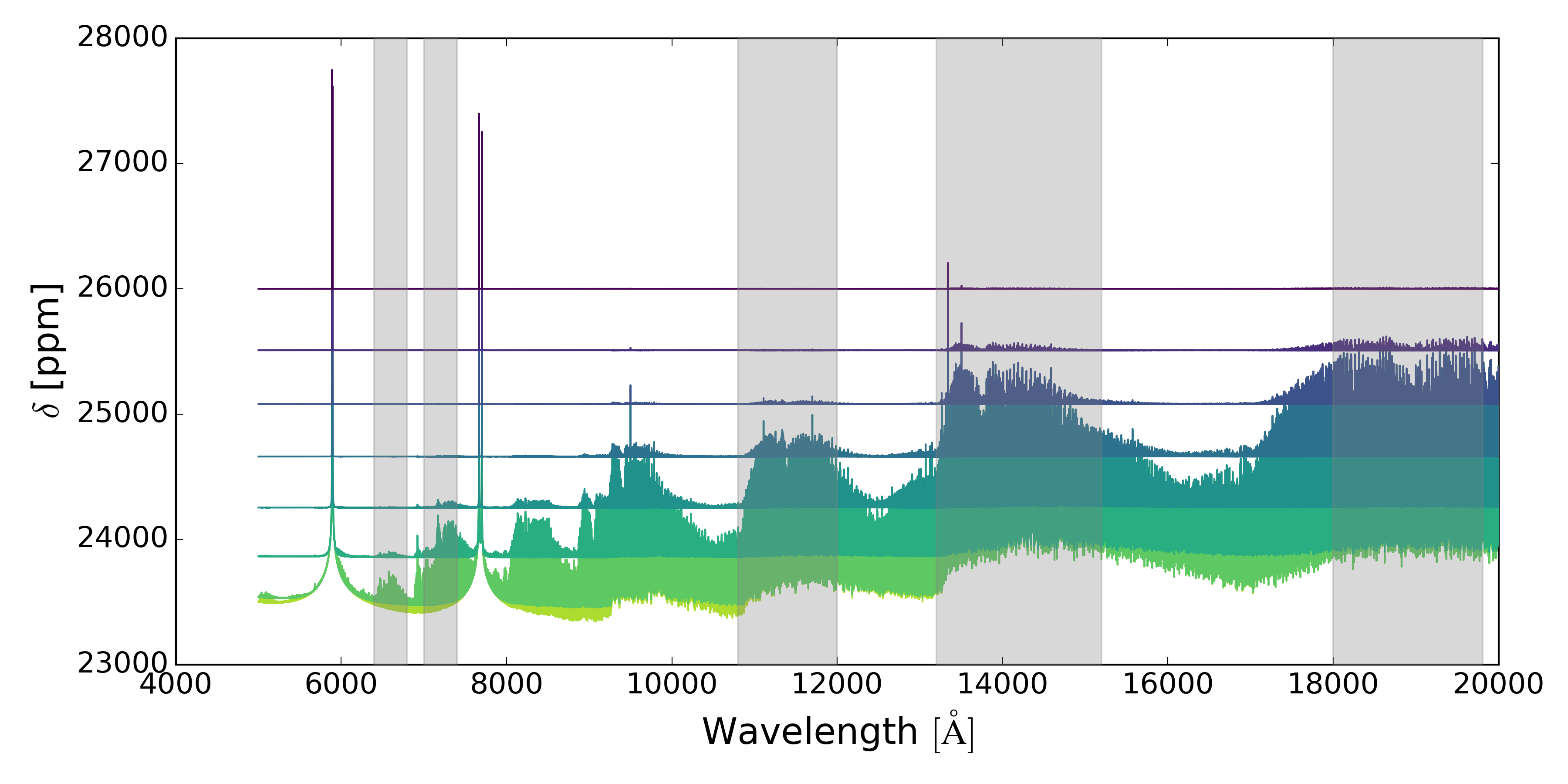}\\
\includegraphics[width=17cm]{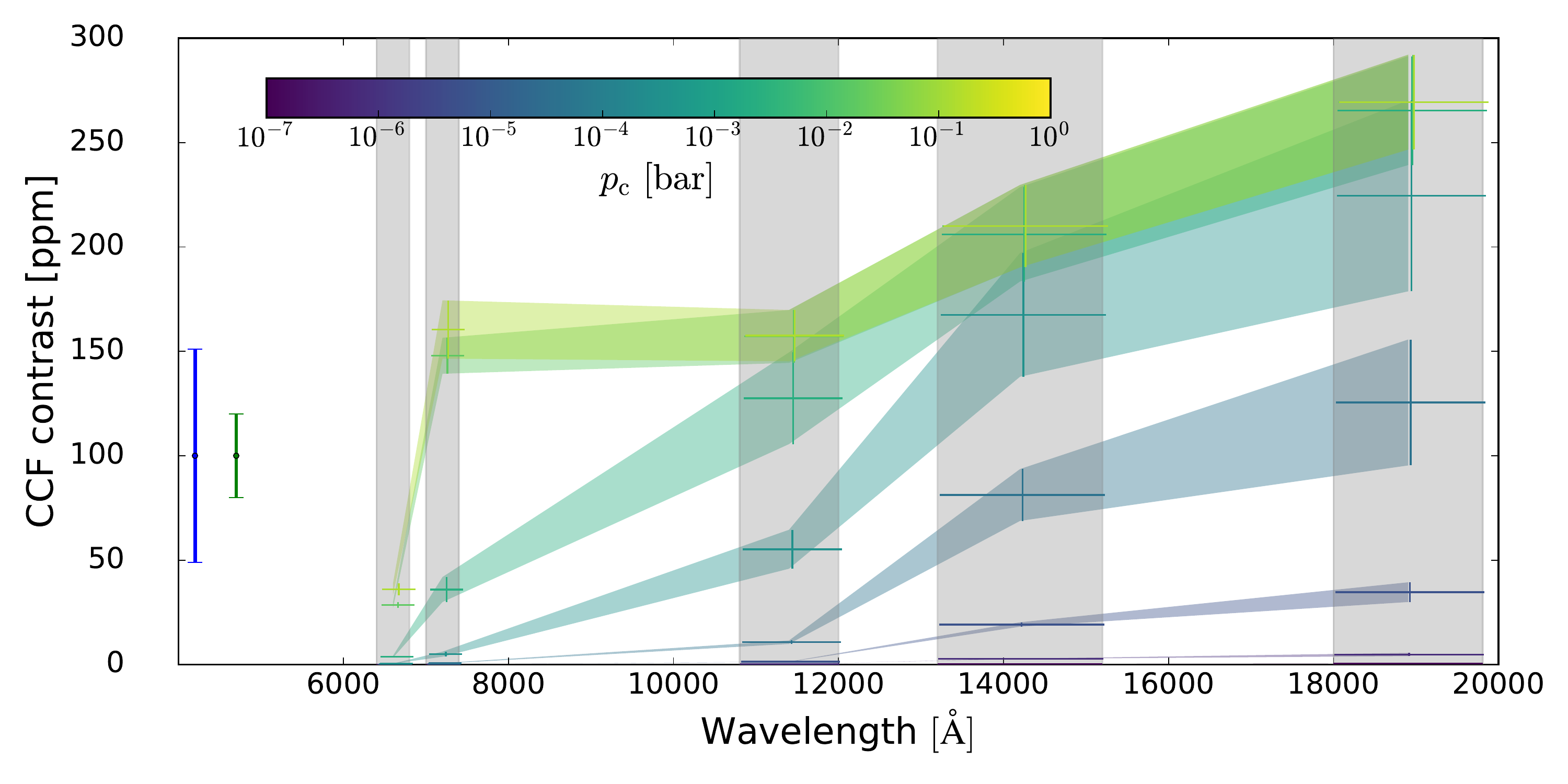}
\caption{Results in the ideal case. \textit{Upper panel:} High resolution models of a hot Jupiter with $T=1\,700~\mathrm{K}$ and different values of $p_\mathrm{c}$, ranging between $10^{-7}~\mathrm{bar}$ and $1~\mathrm{bar}$. Each color corresponds to a different value of $p_\mathrm{c}$, and the lightest colors correspond to condensates deeper in the atmosphere. \textit{Lower panel:} CCF contrast as a function of wavelength. The colors are in correspondence with the models shown in the upper panel. For each value of $p_\mathrm{c}$, the temperature range $1\,200$--$2\,300~\mathrm{K}$ is explored. We indicate with a vertical error bar the induced variation in the contrast of the CCF.  On the left, the 1-pixel, $1\sigma$ precision of HARPS (blue) and ESPRESSO (green) for a single transit are shown). These error bars only account for photon noise, but telluric residuals may introduce additional noise (see Appendix \ref{sec:Detection limits}). The near-infrared bands must be accessed from space or from the stratosphere. Such instruments do not yet exist, and we do not estimate an error bar. \label{Fig:models_clouds}}
\end{figure*}
\section{Results and discussion}
\label{sec:results}
\subsection{Ideal case}
We first discuss the case of solar water abundance with an opaque, gray absorber found at a pressure $p_\mathrm{c}$, representing a population of particles of size $\gtrsim 2~\mu\mathrm{m}$ (see section \ref{sec:template}) in the ideal case (telluric absorption neglected). In Fig. \ref{Fig:CCF_contrast}, we showcase the contrast of the CCF for models at various temperatures as a function of $p_\mathrm{c}$ for two example bands (\espresso{}{}  and \gianothree{}). The following conclusions drawn from this example hold also for the other bands:
\begin{itemize}
\item The CCF of the redder bands have a stronger contrast than the CCF of the bluer bands;
\item The higher in the atmosphere the aerosols are found, the more damped is the signal. A sharp drop in the signal is observed for aerosols higher than a critical pressure;
\item The precise value of this critical pressure depends on which band is considered;
\item The temperature dependence, represented by curves of different colors in Fig. \ref{Fig:CCF_contrast}, is modest.
\end{itemize}
Within our grid, temperature varies by a factor of $\sim2$, but the CCF contrast in each band varies by less than $10\%$ between the coldest ($1\,200~\mathrm{K}$) and the hottest ($2\,300~\mathrm{K}$) model. At first glance, it is surprising that the CCF contrast is not linearly increasing with temperature, since at first order the contrast of every atmospheric signature is proportional to temperature through the scale height. Indeed, the contrast of the sodium and potassium doublets nearly doubles between the hottest and the coldest temperatures considered (Fig. \ref{Fig:temperature_dependence_on_models}, left panel). However, in high-resolution transmission spectroscopy of molecules, the strongly non-linear relation between temperature and population of energetic levels plays an important role, by impacting the relative strength of single molecular lines (see right panel of Fig. \ref{Fig:temperature_dependence_on_models}). The net effect is that, for a hotter atmosphere, the gain in signal due to the increased scale height is partially lost. We emphasize that optimization of detection techniques of molecular signatures through cross-correlation needs to properly account for this factor \citep{Allart2017}.\\
Fig. \ref{Fig:models_clouds} shows the contrast of the CCF of all considered water bands in the ideal case. Here, each color corresponds to a cloud deck at a given pressure, as shown in the upper panel. In the lower panel, the vertical error bar in each band represents the CCF contrast values spun when temperature is varied between $1\,200~\mathrm{K}$ and $2\,300~\mathrm{K}$. The fact that curves corresponding to different values of $p_\mathrm{c}$ do not overlap means it is possible to discriminate among the different cases. The models are best distinguished with a broad wavelength coverage. We indicate as vertical error bars the 1-pixel $1\sigma$ precision of HARPS and ESPRESSO, obtained with the respective Exposure Time Calculators (ETCs) and compatible with \cite{Allart2017}, and valid for HD189733b. The CCF spans about 5 pixels \citep{Allart2017}, implying that the $1\sigma$ photon noise precision on the CCF is even better. However, additional noise is introduced by telluric residuals (see Appendix \ref{sec:Detection limits}). In particular, the technique can only be used for targets with a combination of systemic velocities and BERV such that telluric lines are moved by a few resolution elements; otherwise telluric absorption would contaminate, or even hinder, the planetary signal. The NIR bands investigated in this section are not accessible from the ground for the presence of telluric absorption, and we do not attempt an estimation of the precision of a hypothetical space-borne or stratospheric high-resolution spectrograph (see Appendix \ref{sec:Detection limits} and Sec. \ref{Sec:Methods}). However, it is clear that a HARPS-like precision would suffice to detect the CCF of the NIR water bands at high confidence in the absence of telluric noise.
\subsection{A metric to diagnose aerosols}
\label{sec:delta_C}
\begin{figure}
\resizebox{\hsize}{!}{\includegraphics{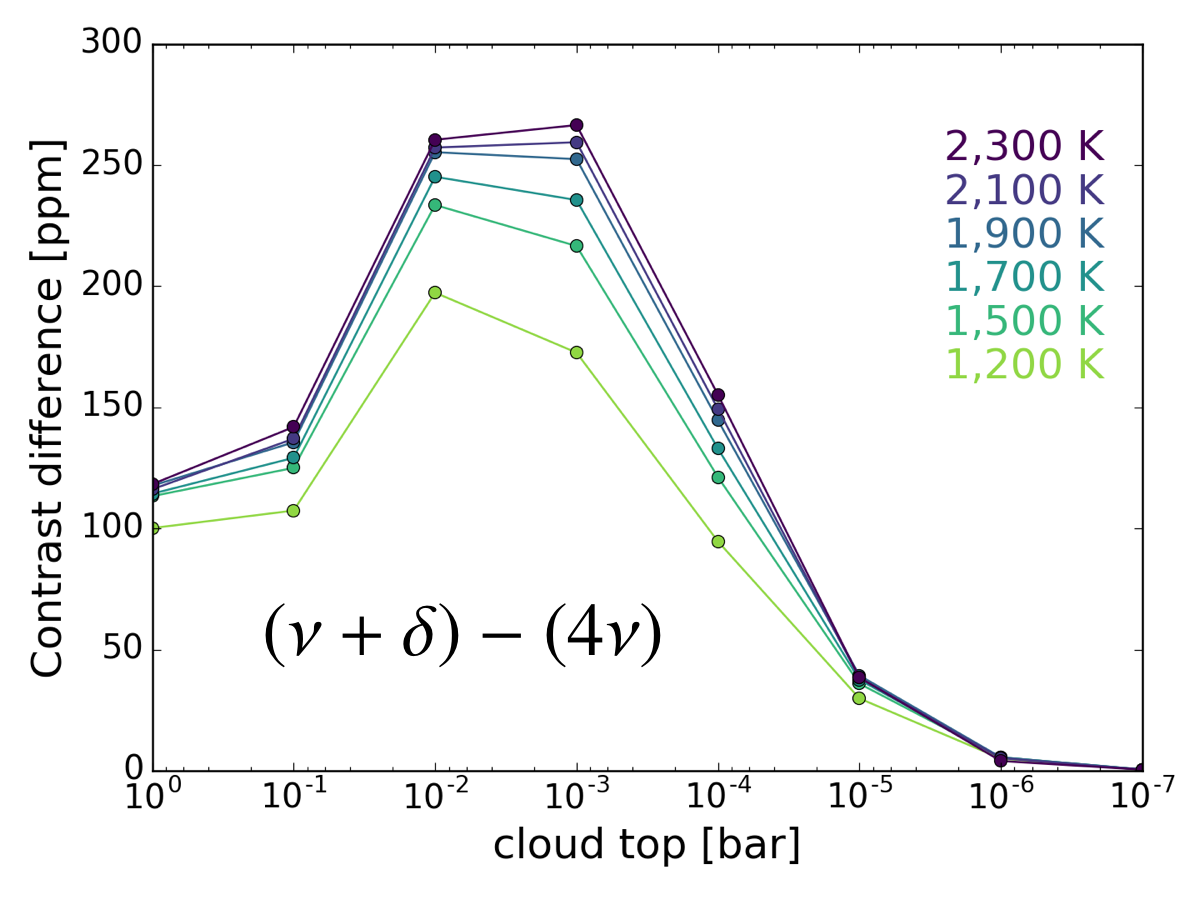}}
\caption{The difference in contrast between the CCF of the \gianothree{} and the \espresso{}{} bands (both shown in Fig. \ref{Fig:CCF_contrast}). The difference is highest when aerosols are found at pressures of $10^{-2}$--$10^{-3}~\mathrm{bar}$. Thus, the measurement of the contrast of both CCFs constrains the presence of aerosols at that height. Note that this pair of water bands is partially sensitive to aerosols lower in the atmosphere as well. \label{Fig:contrast_difference}}
\end{figure}

\begin{figure*}
\includegraphics[width=17cm,trim={2.5cm 0 1.5cm 0},clip]{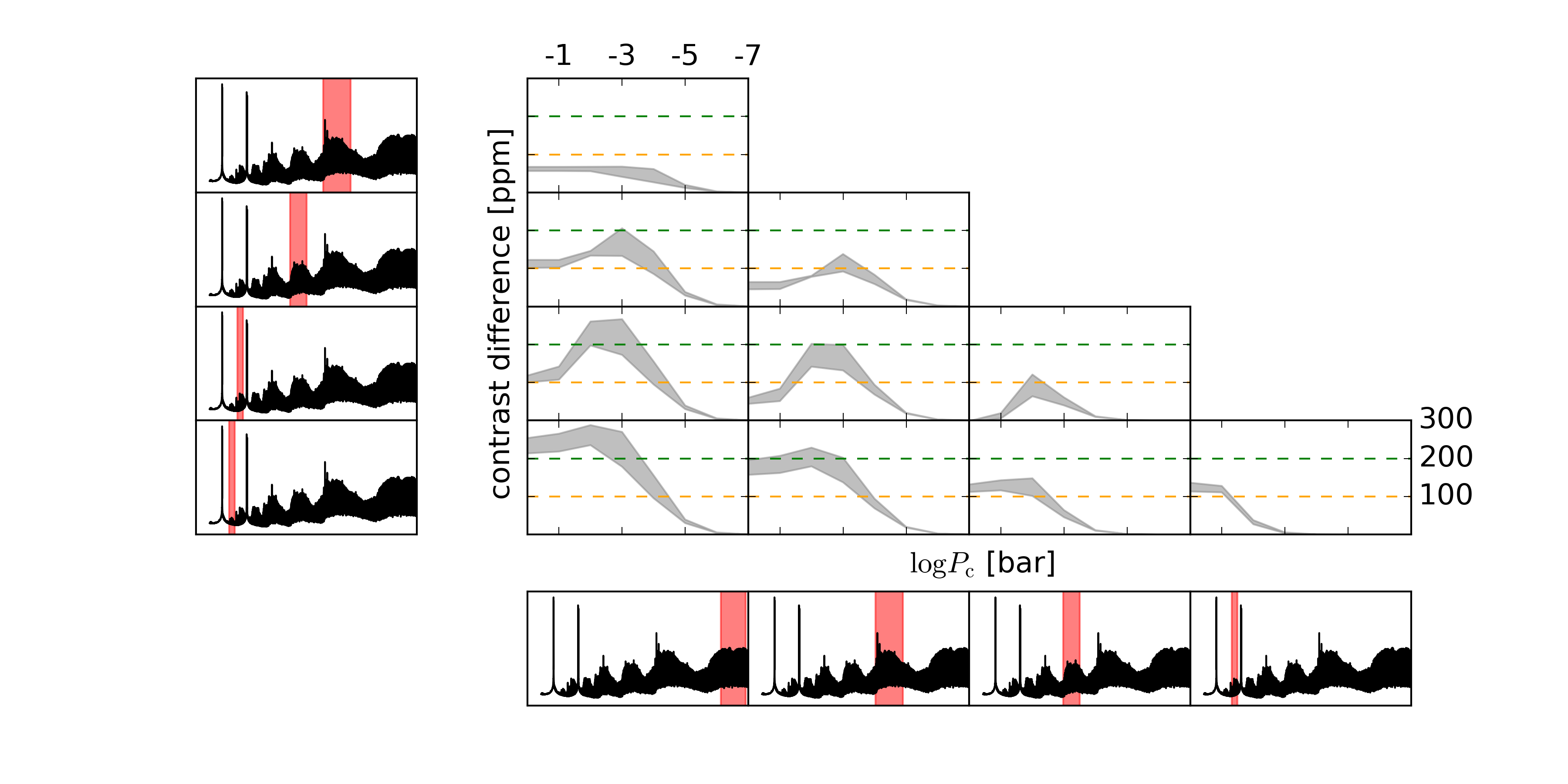}
\caption{Sensitivity of different combinations of water bands to the presence of aerosols. Each panel shows the difference between the band highlighted in the bottom row and the band highlighted in the left column. The x- and y-axes are the same as for Fig. \ref{Fig:contrast_difference}, running between $1~\mathrm{bar}$ and $10^{-7}~\mathrm{bar}$ and between $0~\mathrm{ppm}$ and $300~\mathrm{ppm}$ respectively. Orange and green horizontal dashed lines are put in correspondence of the $100~\mathrm{ppm}$ and the $200~\mathrm{ppm}$ level.
\label{Fig:full_contrats_difference_ideal}}
\end{figure*}

The fact that the critical pressure above which the CCF is damped  differs from band to band is crucial to discriminate aerosol-free and aerosol-rich atmospheres. This is quantified by the difference between the CCF contrast of pairs of bands ($\Delta \mathrm{C}$). In Fig. \ref{Fig:contrast_difference}, we showcase this difference for the bands \gianothree{} and \espresso{}{}. This difference is significantly higher when aerosols are found around $1$--$10~\mathrm{mbar}$, high enough to conceal the \espresso{}{} water band at $7\,200~\mathrm{\AA}$ but not the \gianothree{} water band at $19\,000~\mathrm{\AA}$. The contrast difference $\Delta \mathrm{C}$ between different pairs of bands peaks for different pressures of the aerosols, depending on the value of the critical pressure of both bands. Furthermore, the maximum value of $\Delta \mathrm{C}$ indicates how much the pair of bands discriminates among models. We perform this analysis for every pair of bands, and show the results in Fig. \ref{Fig:full_contrats_difference_ideal}. Overall, combinations of bands covering the optical to the near-infrared can discriminate at the $200~\mathrm{ppm}$ level aerosol decks at $1~\mathrm{bar}<p_\mathrm{c}<0.1~\mathrm{mbar}$, and for the considered instruments. This value of $\Delta \mathrm{C}$ is readily detectable with current or upcoming stabilized spectrographs in the optical, and with a space-borne or stratospheric near-infrared instrument of similar performance. Indeed, by rescaling the precision obtained by \cite{Allart2017} to a CCF built with $800$ lines yields a 1-pixel precision of $62~\mathrm{ppm}$ on the CCF contrast with HARPS, and of $24~\mathrm{ppm}$ with ESPRESSO in a single transit (see also Appendix \ref{sec:Detection limits}). We summarize in table \ref{tab:clouds_sensitivity} the maximum value of $\Delta \mathrm{C}$ and the pressure range where it is achieved for different pairs of water bands observed with HARPS-N, ESPRESSO and GIANO as detailed in Tables \ref{tab:CCF_parameters} and \ref{tab:CCF_parameters_tellurics}. Higher resolving power, especially in the NIR, would provide even higher values of $\Delta \mathrm{C}$.
\begin{table*}
\caption{Typical CCF contrast difference ($\Delta\mathrm{C}$) between pairs of water bands and pressure range where this maximum difference is reached. See also Fig. \ref{Fig:full_contrats_difference_ideal} \label{tab:clouds_sensitivity}}
\centering
\begin{tabular}{lcc}
\hline 
Water bands & Cloud pressure $[\mathrm{bar}]$& $\Delta\mathrm{C}$ $[\mathrm{ppm}]$ \\ 
\hline 
\\
\gianothree{}-\gianotwo{} & $1$--$10^{-4}$ & $50$\\ 
\gianothree{}-\gianoone{} & $10^{-2}$--$10^{-4}$ & $150$--$200$\\ 
\gianothree{}-\espresso{}{} & $10^{-2}$--$10^{-3}$ & $200$\\ 
\gianothree{}-\harps{} & $1$--$10^{-3}$ & $250$\\ 
\gianotwo{}-\gianoone{} & $10^{-3}$ & $100$\\ 
\gianotwo{}-\espresso{}{} & $10^{-2}$--$10^{-3}$ & $150$--$200$\\ 
\gianotwo{}-\harps{} &  $1$--$10^{-3}$ & $150$--$200$\\ 
\gianoone{}-\harps{} & $1$--$10^{-2}$ & $120$\\ 
\gianoone{}-\espresso{}{} & $10^{-2}$ & $100$\\ 
\espresso{}{}-\harps{} & $1$--$10^{-1}$ & $100$\\ 
\\
\hline 
\end{tabular}
\end{table*}
\noindent \subsection{Telluric contamination}
\begin{figure*}
\includegraphics[width=17cm,trim={2.5cm 0  1.5cm 0},clip]{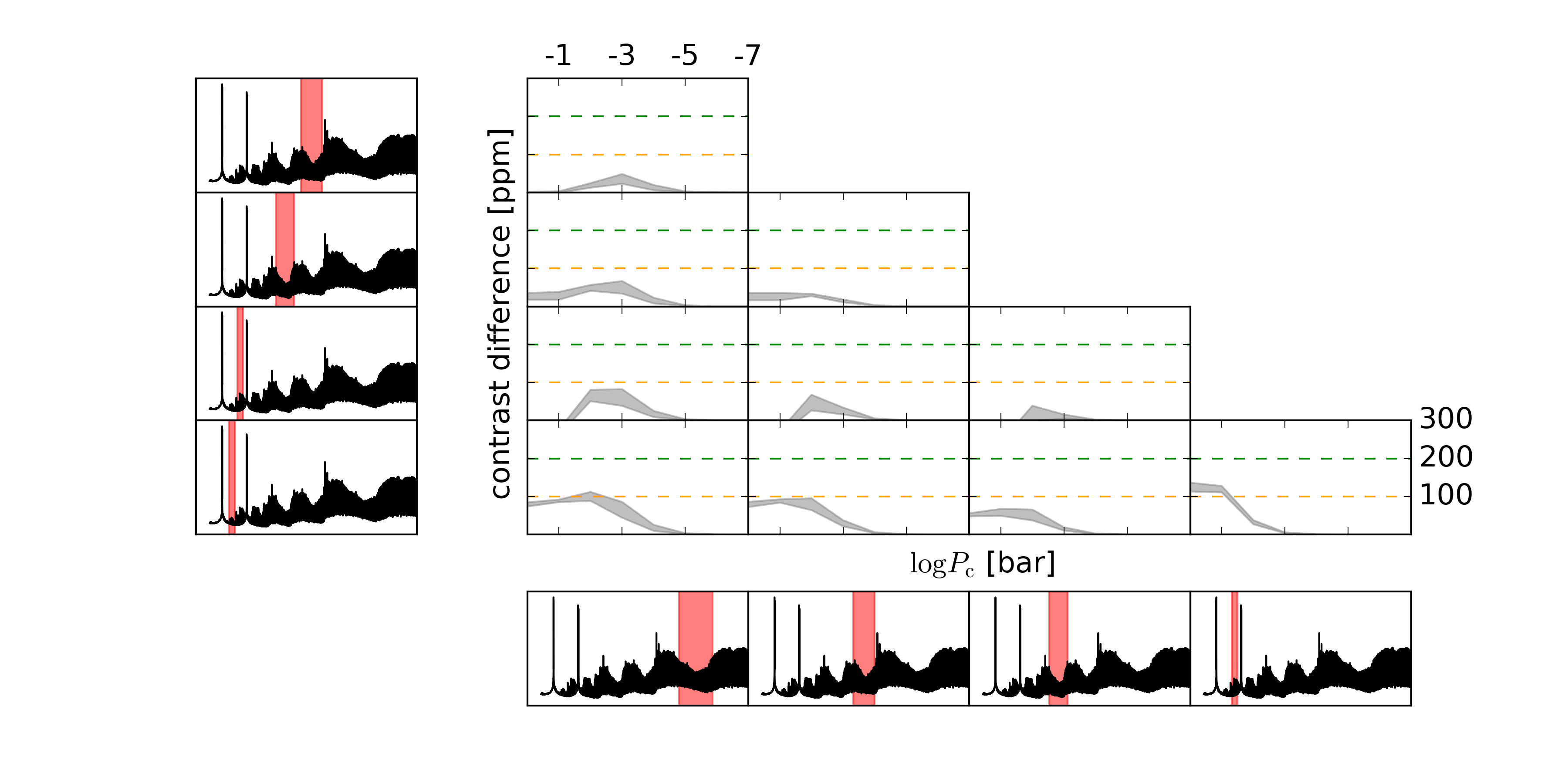}
\caption{Sensitivity to the presence of aerosols of different combinations of bands that are accessible from the ground even in the presence of tellurics. Each panel shows the difference between the band highlighted in the bottom row and the band highlighted in the left column. The x- and y-axes are the same as for Fig. \ref{Fig:contrast_difference}, running between $1~\mathrm{bar}$ and $10^{-7}~\mathrm{bar}$ and between $0~\mathrm{ppm}$ and $300~\mathrm{ppm}$ respectively. Orange and green horizontal dashed lines are put in correspondence of the $100~\mathrm{ppm}$ and the $200~\mathrm{ppm}$ level. Note that these values are arbitrary. Scatterers up to the $0.1~\mathrm{mbar}$ level produce values of $\Delta \mathrm{C}$ up to $100~\mathrm{ppm}$.
\label{Fig:full_contrats_difference_tellurics}}
\end{figure*}

We repeated the analysis focusing on the infrared transparency windows of Earth. These coincide with the transparency windows in the exoplanet. In transmission geometry, the loss in absorption is not so abrupt because the stellar light-rays cross transversely the atmosphere. The net result is that the analysis is still sensitive to the presence of aerosols, but at the reduced level of $\sim 100~\mathrm{ppm}$ (see Table \ref{tab:clouds_sensitivity_tellurics} and Fig. \ref{Fig:full_contrats_difference_tellurics}). This is likely already enough to detect the presence of aerosols with the simulated GIARPS and ESPRESSO observations (see Table \ref{tab:errors_on_CCF}). This is demonstrated by adding to our simulated observations the impact of telluric absorption, as described in Appendix \ref{sec:Detection limits}. We account for both the increase in the photon noise due to the reduced photon count in the core of telluric water lines, and for residuals of an imperfect telluric correction.\\
Moreover, at least two strategies to further increase the signal are possible:
\begin{enumerate}
\item We did not fully explore the wavelength range. CARMENES is sensitive to two other water bands not considered in this work, that are less intense but much less contaminated by telluric lines, thus likely accessible from the ground; GIANO is sensitive also to the K band, not considered here;
\item We simulated GIANO observations in the NIR. GIANO has a resolving power of $\sim50\,000$, however instruments at higher resolving power are available (CARMENES) and several others will be in the near future (NIRPS, SPIRou, CRIRES+, ...). A higher resolving power has the advantage that the CCF contrast is higher, because of the narrower LSF and because less molecular lines are blended in the spectrum (see Fig. \ref{Fig:resolving_power_effect}).\\
\end{enumerate}
\begin{table*}
\caption{Typical CCF contrast difference ($\Delta\mathrm{C}$) between pairs of water bands and pressure range where this maximum difference is reached when considering regions where telluric correction is possible. See also Fig. \ref{Fig:full_contrats_difference_tellurics}.
\label{tab:clouds_sensitivity_tellurics}
}
\centering
\begin{tabular}{lcc}
\hline 
Water bands & Cloud pressure $[\mathrm{bar}]$& $\Delta \mathrm{C}$ $[\mathrm{ppm}]$ \\ 
\hline 
\\
H band-J band & $10^{-3}$ & $50$\\ 
H band-Y band & $10^{-2}$--$10^{-3}$ & $50$\\ 
H band-\espresso{}{} & $10^{-2}$--$10^{-3}$ & $50$--$100$\\ 
H band-\harps{} & $1$--$10^{-2}$ & $100$\\ 
J band-Y band & $1$--$10^{-2}$ & $25$\\ 
J band-\espresso{}{} & $10^{-2}$--$10^{-3}$ & $50$--$25$\\ 
J band-\harps{} &  $1$--$10^{-2}$ & $100$\\ 
Y band-\harps{} & $1$--$10^{-2}$ & $50$\\ 
Y band-\espresso{}{} & $10^{-2}$ & $<50$\\ 
\espresso{}-\harps{} & $1$--$10^{-1}$ & $100$\\ 
\\
\hline 
\end{tabular}
\end{table*}
\subsection{Non-gray aerosols}
\label{sec: non-gray}
The assumption that aerosols have a gray scattering cross-section is only valid for particles of size larger than about $2~\mu\mathrm{m}$, and up to $\lambda\sim3~\mu\mathrm{m}$. We discuss the more general scenario of non-gray aerosols, without performing full simulations that would require an implementation of the Mie theory of scattering.\\
At wavelengths longer than $3~\mu\mathrm{m}$, distinctive absorption features of aerosols species such as water or silicates become prominent in transmission spectra \citep{DeMooij2013, Molliere2017}. These relatively narrow features uniquely identify the composition of aerosols, making them probes of the chemistry of exoplanets. However, at wavelengths longer than $5~\mu\mathrm{m}$, where the majority of such features are located, the sky background is too bright for ground-based spectrographs to be effective \citep{Snellen2013}.\\
In the wavelength range $<2~\mu\mathrm{m}$ that we considered in this paper, the Rayleigh limit of Mie scattering from ``small'' particles with sizes typically smaller than about $1~\mu\mathrm{m}$ results in a cross-section of aerosols that decreases with wavelength with a typical slope, if they are present (e.g. \citealt{Pinhas2017}; the precise particle size depends also on the wavelength of incident light and is smaller in the optical).  Several hot Jupiters host aerosols of this kind \citep{Sing2016, Barstow2017}. Compared to the gray aerosols considered heretofore, this scenario exacerbates the difference in contrast between the optical water bands, even more affected by scattering from this kind of aerosols, and the near-IR water bands. $\Delta\mathrm{C}$ is thus increased, and aerosols are readily detected also in this case. We tested this by simulating an atmosphere compatible with HD189733b, which hosts a marked optical slope attributed to this kind of particles \citep{Pont2013, Pino2018}. In this case $\Delta \mathrm{C}$ reaches up to $100~\mathrm{ppm}$, and aerosols are thus promptly detected.\\
The chromatic dependence of the cross-section of small aerosols makes it more challenging to locate their height in the atmosphere. Yet, its slope and uniformity indicate the distribution of the sizes of the scattering particles and their composition \citep{Berta2012, Pont2013, Wakeford2015, Pinhas2017}. An extension of the technique that we presented has thus the potential to characterize the population of aerosols in the atmosphere. However, a detailed assessment requires detailed modelling of Mie scattering by aerosols, and is out of the scope of this paper.\\
\cite{Snellen2004}, \cite{DiGloria2015} and \cite{Heng2016} presented alternative techniques that are able to constrain the presence of chromatic scattering by aerosols through high-resolution spectrographs of the kind we considered. If its presence can be excluded using such techniques, $\Delta\mathrm{C}$ can be used to infer the altitude of aerosols as discussed in Sec. \ref{sec:delta_C}.
\noindent \subsection{Water abundance}
\begin{figure}
\resizebox{\hsize}{!}{\includegraphics{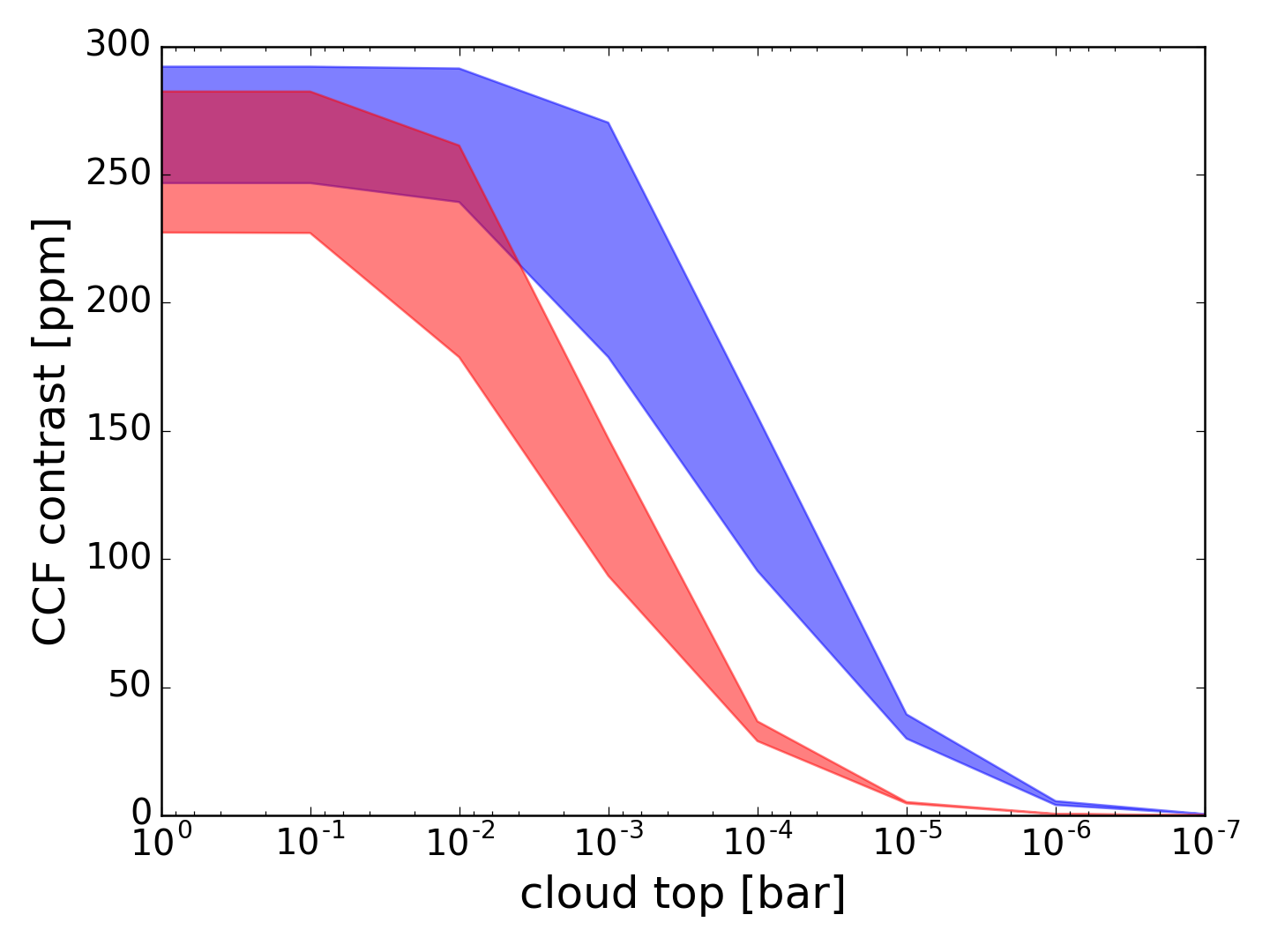}}
\caption{\label{Fig:compare_solar_subsolar} Contrast of the CCF of the \gianothree{} band, observable with GIANO, as a function of altitude of aerosols, expressed as pressure of the cloud top. The shadowed area is spanned by the models when the temperature is varied between $1\,200$ and $2\,300~\mathrm{K}$. A sharp decrease in the contrast of the CCF is caused by aerosols higher than a critical pressure. For solar water abundance, this threshold is located at about $10^{-3}~\mathrm{bar}$ (blue curve). For 0.1 times solar abundance, this threshold is located at about $10^{-2}~\mathrm{bar}$ (red curve), thus 10 times deeper in the atmosphere.}
\end{figure}

Throughout the paper we assumed a solar VMR for water. This is broadly supported by equilibrium chemistry calculations over a broad range of temperatures (e.g. \citealt{Sharp2007, Miguel2014}) and by current observational evidence (e.g. \citealt{Crossfield2015, Todorov2016}). However, determining the absolute abundance of water from observations is challenging (see e.g. \citealt{Kreidberg2014, Wakeford2018}). We first discuss the observational strategy to measure the water mixing ratio in transmission and its limitations. Then, we show that an uncertainty in the VMR of water directly translates in an uncertainty in the altitude of aerosols inferred under the assumption of a gray aerosol deck (see Sec. \ref{sec:delta_C}). A similar impact on the inferred properties of aerosols (particle size distribution, composition, etc.) is expected in the more general case of chromatic scattering by aerosols (see Sec. \ref{sec: non-gray}).
\paragraph{Measuring the VMR of atomic and molecular species with ground-based, high-resolution transmission spectroscopy:}
High-resolution spectra from the ground can only provide measurements of the difference between the transit depths at different wavelengths. We now consider two atomic/molecular atmospheric species 1 and 2, that are the dominating source of opacity at wavelengths $\lambda_1$ and $\lambda_2$, respectively. \cite{Etangs2008, Desert2009} first showed how to relate the difference between the transit depths at these two wavelengths to the abundance ratio of the atmospheric species. The same result can be obtained by using Eq. (12) by \cite{Heng2017}, valid in the isothermal, isobaric case. Indeed, by dividing by the stellar radius and differentiating this equation, the difference in the transit depth at the two wavelengths can be related to the volume mixing ratio and the cross-section of the dominating spectral feature at the two wavelengths as
\begin{equation}
\label{Eq:diff_tr_depth}
\dfrac{R(\lambda_2)}{R_\star} - \dfrac{R(\lambda_1)}{R_\star} = \dfrac{H}{R_\star}\ln \left( \dfrac{\mathrm{VMR_2}\sigma_2}{\mathrm{VMR_1}\sigma_1}\right)\ .
\end{equation}
The scale height $H$ can be inferred by comparing the observed planet radius at different wavelengths where the cross-section is known \citep{Benneke2012}. At high resolution, this is possible by using multiple bands of the same molecule where aerosols have a small impact. Unfortunately water is not well suited to the task from the ground, since the contrast of the CCF in the Y, J and H transparency windows is impacted by relatively low aerosols at the $10^{-1}~\mathrm{bar}$ level. Other molecules such as methane and titanium monoxide, or atomic species such as the alkali doublets \citep{Heng2016}, may be better suited, since they are less impacted by telluric absorption.\\
Without tellurics contamination, e.g. from a hypothetical space-borne or stratospheric high-resolution spectrograph, the water bands \gianothree{} and \gianotwo{} are ideally suited. Indeed by rearranging Eq. \ref{Eq:diff_tr_depth} and evaluating it in the \gianothree{} and \gianotwo{} water bands we can isolate the scale height:
\begin{equation}
\dfrac{H}{R_\star} = \dfrac{R_{\left(\nu+\delta\right)} -R_{\left(2\nu\right)}}{R_\star \ln \left(\dfrac{\left<\sigma_{\mathrm{H_2O}} \right>_{\left(\nu+\delta\right)}}{\left<\sigma_{\mathrm{H_2O}} \right>_{\left(2\nu\right)}}\right)}\ .
\end{equation}
From Fig. \ref{Fig:full_contrats_difference_ideal}, we see that the contrast difference between the two bands is constant for aerosols up to $10^{-4}~\mathrm{bar}$. In this region, the contrast difference is set by the intrinsic difference in contrast between the bands, and is thus a probe of scale height. Farther in the infrared, stronger molecular bands may be even better suited for the task.\\
Besides knowing the scale height, a second step is necessary to measure the volume mixing ratio $\mathrm{VMR_2}$ of a given constituent using Eq. \ref{Eq:diff_tr_depth}. Indeed, we need to compare the transit depth in a wavelength region where it is the dominating source of opacity to a region where the dominating source of opacity has a known $\mathrm{VMR}_1$. Then, we can invert Eq. \ref{Eq:diff_tr_depth} to get $\mathrm{VMR_2}$.
\paragraph{Measuring the VMR of atomic and molecular species with space-borne spectroscopy, photometry or multi-object spectroscopy:}
The operations described in the previous paragraph can be performed without directly measuring the transit depth but only comparing excesses in different bands. They are thus possible with ground-based high-resolution facilities. Yet, an hypothetical space-borne or stratospheric high-resolution spectrograph may be able to directly measure the transit depth, as the effect of Earth atmosphere is absent or severely limited. Already now, the transit depth can be measured from space, or with photometric and multi-object spectroscopic techniques from the ground, albeit at lower resolution (up to $R\sim 10^3$, too low to apply the CCF technique).\\
However, even a direct measurement of the transit depth within a molecular band is not sufficient to infer its VMR. Indeed, the pressure-radius\footnote{Either of the two can be arbitrarily fixed, and the other becomes a free parameter.} $p_\mathrm{ref}(r_\mathrm{ref})$ at an arbitrary reference level must be known. This can be obtained by directly measuring the transit depth in an absorption feature generated by a species of known VMR \citep{Etangs2008}.\\
\cite{Etangs2008} proposed  Rayleigh scattering by $\mathrm{H_2}$ as such a comparison feature. However, the task of detecting the absorption feature of molecular hydrogen is hindered in the presence of Rayleigh scattering by aerosols that may contaminate the optical region of the spectrum. As discussed in  Sec. \ref{sec: non-gray} measuring the slope of the optical transmission spectrum and its uniformity provides hints on whether aerosols of small size are present or not in the atmosphere.
\paragraph{Effect of an uncertain water VMR:}
It is thus worth getting an idea of the effect of an uncertain water abundance on our conclusions. Analytical formulae provide a first insight.
\cite{Etangs2008} defined the equivalent altitude $z_{\mathrm{eq}}$ by the altitude in the atmosphere such that a sharp occulting disk of radius equal to the sum of the planetary radius and $z_{\mathrm{eq}}$ produces the same absorption depth as the planet and its translucent atmosphere. We can reformulate the concept as an equivalent pressure $p_{\mathrm{eq}}$ by
\begin{equation}
p_\mathrm{eq}=p_{\mathrm{ref}}\exp\left( z_\mathrm{eq}/H \right)\,,
\end{equation}
where $p_{\mathrm{ref}}$ is the pressure at the arbitrarily chosen reference level. Within a water band $n$, water is the dominant source of opacity, thus we can write
\begin{equation}
\label{Eq:equivalent pressure}
p_\mathrm{eq,\,n}=\dfrac{\mathrm{VMR}_\mathrm{H_2O}<\sigma_{\mathrm{H_2O}}>_n p^2_{\mathrm{ref}}}{\tau_{\mathrm{eq}}}\sqrt{\dfrac{2\pi r_p}{k_B T\mu g}}\,,
\end{equation}
where $<\sigma_{\mathrm{H_2O}}>_n$ is the average cross-section of the water lines within the $n-$th molecular band, $\tau_{\mathrm{eq}}\sim 0.56$ is the equivalent optical depth \citep{Etangs2008}, $T$ is the atmospheric temperature, $\mu$ is the mean molecular weight of the atmospheric constituents and $g$ is the surface gravity.\\
At first order, to determine if a water band is occulted or not by aerosols, we can compare its equivalent pressure with the pressure where aerosols are located. Thus, the drop in the contrast of the CCF shown in Fig. \ref{Fig:CCF_contrast} is located around the equivalent pressure within the considered water band.\\
To qualitatively understand the effect of the variation of $\mathrm{VMR_{H_2O}}$, it is sufficient to note that, from Eq. \ref{Eq:equivalent pressure}, it follows that
\begin{equation}
\label{Eq:equivalent_pressure_and_abundance}
\dfrac{p_\mathrm{eq,\,n,\, \odot}}{p_\mathrm{eq,\,n}}=\dfrac{10^{-3}}{\chi_{\mathrm{H_2O}}}\,,
\end{equation}
where $p_\mathrm{eq,\,n,\, \odot}$ is the equivalent pressure when water is present in the atmosphere at solar abundance.\\
We tested numerically our intuition, by repeating the full analysis with $\mathrm{VMR_{H_2O}}=10^{-4}$ (about 0.1 the solar value). In Fig. \ref{Fig:compare_solar_subsolar} we compare the contrast of the CCF of the \gianothree{} band for $\mathrm{VMR_{H_2O}}=10^{-3}$ (blue) and $\mathrm{VMR_{H_2O}}=10^{-4}$ (red). When $\mathrm{VMR_{H_2O}}$ is decreased by a factor of 10, aerosols located at pressures of a factor of 10 higher, thus deeper in the atmosphere, are sufficient to occult the water band, confirming Eq. \ref{Eq:equivalent_pressure_and_abundance}.\\
The conclusion that high-resolution transmission spectroscopy of water over a broad wavelength range can be used infer the presence and constrain the properties of aerosols, such as their altitude in the simulated case of gray scattering, still holds. However, an unknown water abundance introduces an uncertainty in the aerosols properties, which must be accounted for.\\

\section{Conclusions and future prospects}
High-resolution, ground-based transmission spectra are obtained using double-normalized data reduction techniques. These techniques remove broad band features such as those due to aerosols. Here, we demonstrated that the contrast difference of the CCF of different absorption bands of water ($\Delta \mathrm{C}$), accessible with this kind of measurements, can still be used as a probe of the presence of aerosols. A key factor for the success of this technique is having a broad wavelength coverage, which is provided by current and forthcoming instrumentation (see Sec. \ref{sec: intro}).\\
We limited our analysis to the contrast of the CCF, and thus did not consider the full information encoded in the CCF. For example, the broadening of the CCF is a function of temperature, rotation and atmospheric dynamics. We thus plan to extend our technique to consider the full CCF as a probe of the conditions in the atmosphere. Furthermore, the technique can be extended to other molecular species. Methane is a good candidate as indicator of aerosol-rich atmospheres, as it possess a wealth of absorption bands (that in general do not coincide with the strong water and oxygen telluric absorption bands) that span a broad wavelength range evenly. It can be expected in planets colder than the one simulated here, thus extending the range of cases where our technique can be applied. CO also possesses two absorption bands in the J and K bands, at $1.56~\mu\mathrm{m}$ and $2.34~\mu\mathrm{m}$, potentially observable in transmission (e.g. \citealt{Brogi2016}) whose intensity can in principle be compared. They have the advantage that CO is not a strong telluric absorber and of being in the transparency window of Earth; however, with two bands, less information can be extracted by using our technique. On the hot end of the hot Jupiter spectrum ($T>2\,000~\mathrm{K}$), titanium monoxide (TiO) is a good candidate for the same reasons.\\
\cite{Hoeijmakers2015} and \cite{Brogi2017} demonstrated that, at high-resolution, inaccuracies in molecular line lists propagate to models, introducing uncertainties both in line position and strength (see also a discussion in \citealt{Pino2018}). While this aspect needs to be thoroughly tested, progress is being made. The ExoMol line list \citep{Tennyson2016}, that contains billions of lines, is computationally challenging to use but may possess the required accuracy for this kind of studies \citep{Yurchenko2008, Yurchenko2011}.\\
The presence of multiple molecular species with overlapping bands and blended lines might also change the continuum level, and possibly the value of $\Delta\mathrm{C}$. However, in the wavelength range considered ($\lambda<2~\mu\mathrm{m}$), only water and titanium monoxide were detected. While other constituents are most likely present, they are undetected, indicating that their signals must be less than the precision of current instrumentation. For this reason, their effect on the continuum should be negligible compared to that of aerosols. Furthermore, $\mathrm{TiO}$ is only expected in the hottest atmospheres ($T_\mathrm{eq}\gtrsim2\,000~\mathrm{K}$). Our technique can thus be safely applied to hot Jupiters with temperatures between about $1\,000$ and $2\,000~\mathrm{K}$, but for hotter planets care must be taken in the interpretation of $\mathrm{\Delta C}$, and TiO included in the analysis. However, fewer aerosols are expected on this kind of planets (e.g. \citealt{Heng2016, Stevenson2016}). In addition, the TiO line list is unreliable at the shortest wavelengths considered in this paper \citep{Hoeijmakers2015}, thus we do not simulate this case.\\
An extension of the technique to wavelengths $\lambda>2~\mu\mathrm{m}$ and to include chromatic scattering by aerosols is warranted, as it may enable the characterization of the composition and size distribution of aerosol particles. Such an extension requires modeling Mie scattering from aerosols and modeling molecular spectral features from multiple species at high-resolution.\\

\appendix
\section{Estimate of photon and telluric noise in CCF observations}
\label{sec:Detection limits}
A complete assessment of noise sources in real observations of the water CCF is out of the scope of this paper. We do not simulate systematic instrumental effects, since they are different for different instruments, and only partially understood. We stress that having a single instrument able to cover multiple bands such as CARMENES, or a combination of instruments sharing the optics such as HARPS-N + GIANO or HARPS + NIRPS, is generally preferable to combining non-simultaneous observations from different instruments. Instead, we provide estimates of the impact of photon noise and imperfect telluric removal, which are shared by all observations provided by ground-based instrumentation.\\

\subsection{Telluric contamination}
\begin{figure*}
\resizebox{\hsize}{!}{\includegraphics{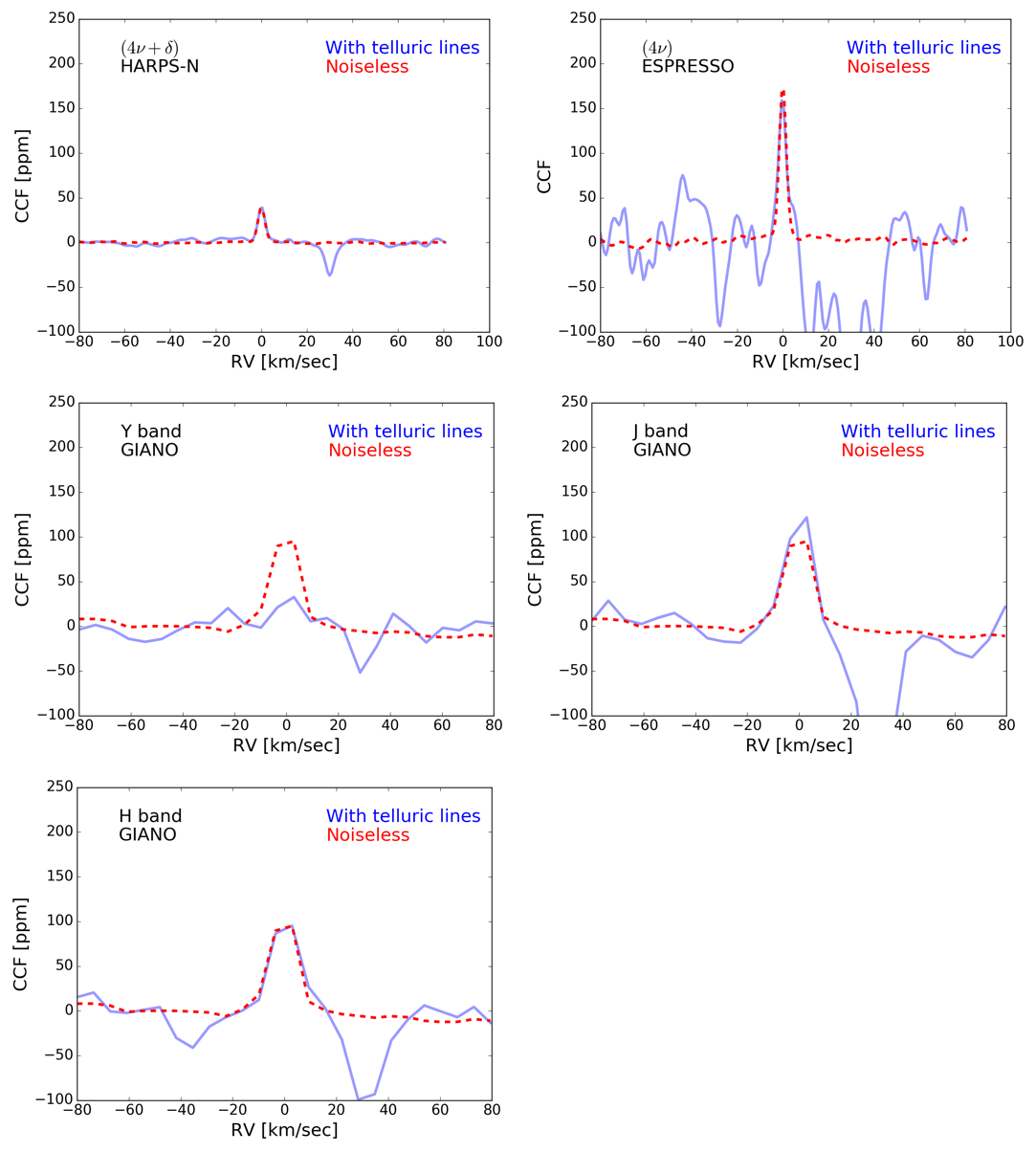}}
\caption{Effect of telluric contamination on observations of an aerosol-free HD189733b with HARPS-N (upper-left panel), ESPRESSO (upper-right panel) and GIANO (lower-left panel). Red solid line: contamination-free transmission spectrum. Blue solid line: transmission spectrum in the presence of telluric residualso consistent with \cite{Allart2017}. If the systemic velocity and the Barycentric Earth Radial Velocity combine such that the exoplanetary signal is shifted away from the telluric residual (here: $30~\mathrm{km\,sec^{-1}}$), the signal is recovered in the \harps{}, \espresso{}, J and H bands. In the Y band, a more careful analysis than what performed here is possible, and may lead to a detection (see main text). In the \gianoone{}, \gianotwo{} and \gianothree{} bands, water absorbs all the incident light making detection impossible from the ground.\label{Fig:telluric_CCF}}
\end{figure*}

\begin{figure*}
\resizebox{\hsize}{!}{ \includegraphics{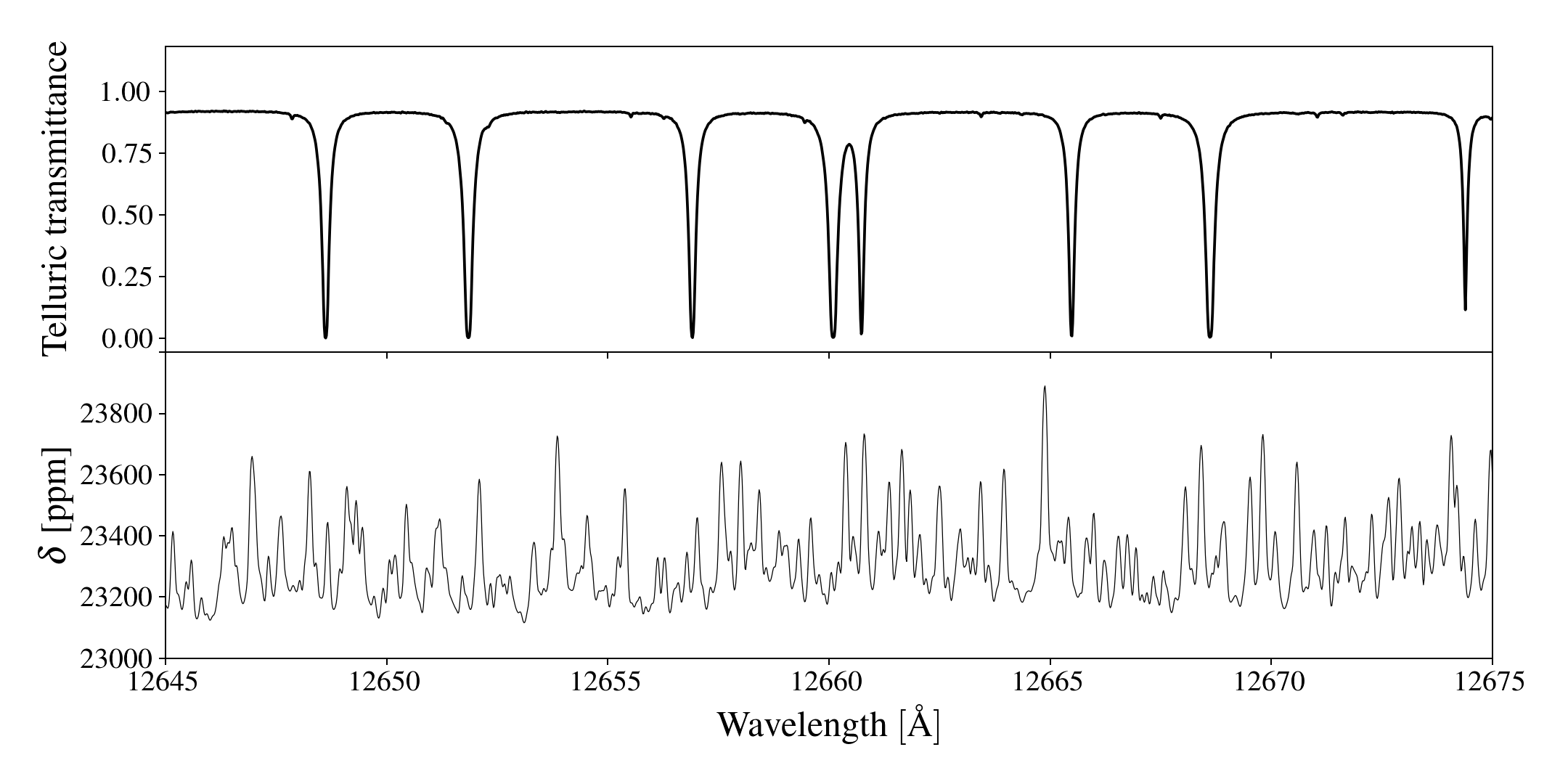}}
\caption{The properties of water and molecular oxygen absorption are very different because of the different structure of the molecules. \textit{Upper panel:} transmittance of Earth in the J band. The deep telluric lines are from molecular oxygen. They are well separated, and the transmittance approaches 1 between them. \textit{Lower panel:} an aerosol-free model of HD189733b in the same region. Water generates a dense forest of lines, such that the exoplanetary signal can be recovered from the ground from in-between the deep oxygen telluric lines.}\label{Fig:zoom_oxygen_band}
\end{figure*}

In the next future, the kind of observations that we simulated will be carried out only from the ground. We thus estimate the impact of imperfect telluric correction on the simulated observations.\\
We adopt a telluric transmittance spectrum $T_1(\lambda)$ based on observations of the Sun \citep{Hinkle2003}. We scale the telluric spectrum, given at airmass 1, at a generic airmass following \cite{Vidal-Madjar2010}:
\begin{equation}
T_a(\lambda)=(T_1)^a(\lambda)\ ,
\end{equation}
where $a$ is the target airmass. A transmission spectrum is built by taking the ratio of in- and out-of-transit spectra. We assume that these are observed at typical airmasses 1 and 1.2, respectively. The exact value chosen is not critical. We inject residuals of imperfect telluric correction in our transmission spectra according to:   
\begin{equation}
\begin{split}
& \delta_\mathrm{tell}(\lambda) = \\
& = 1- \left\{1-\mathcal{C}\left[1-\dfrac{T_1}{T_{1.2}}\cdot
\lambda\left(1+\dfrac{\mathrm{BERV}}{c}\right)\left(1+\dfrac{v_{\mathrm{syst}}}{c}\right)\right]\right\}\cdot \\
& \cdot(1-\delta(\lambda))\ ,
\end{split}
\end{equation}
where $\delta(\lambda)$ is the telluric-free model and the telluric spectrum is injected to account for the non-zero systemic velocity of the target star $v_{\mathrm{syst}}$ and the BERV (Barycentric Earth Radial Velocity). For a typical case, we fix $v_{\mathrm{syst}}=-2.27\mathrm{km\ sec^{-1}}$ as for HD189733b \citep{Boisse2009} and take a BERV of about $30~\mathrm{km~s^{-1}}$. The constant $\mathcal{C}$ ranges between 0 and 1, and represents the quality of the telluric correction. For $\mathcal{C}=0$ the correction is ideal, i.e. there are no telluric residuals, for $\mathcal{C}=1$ there is no correction. We tune $\mathcal{C}$ in order to have the telluric residual in the simulated HARPS CCF within the noise level on the CCF obtained by \cite{Allart2017} on HD189733b. This is obtained by setting $\mathcal{C}=2\%$, and constitutes a worst case scenario for the quality of telluric correction with their technique.\\
It is safe to assume that with ESPRESSO a better quality telluric correction level will be achieved. On the other hand, GIANO data will always be obtained simultaneously to HARPS-N. With the exception of $\mathrm{OH}$ emission features, the strongest telluric bands in the NIR are due to the same species that produce them in the optical. Thus, the model transmission spectrum obtained in the optical can be extended to the GIANO wavelength range, and refined with the additional information, to correct for most of the telluric features. In the NIR range other techniques that exploit the change in the radial velocity of the planet are available (e.g. \citealt{Brogi2018}). Here, we do not simulate them.\\
 In the \gianoone{}, \gianotwo{} and \gianothree{} water bands telluric lines saturate, leaving no flux for ground-based observations (see Fig. \ref{Fig:CCF_parameters}). However, Fig. \ref{Fig:telluric_CCF} shows that the CCF signal is recovered in the \harps{}, \espresso{}, J and H bands even in the presence of telluric residuals. Note that the \espresso{} band is located within a stronger telluric water band, which results in stronger residuals; however, as already mentioned, our estimate of the telluric correction in this region is likely pessimistic since based on HARPS data, of lower quality with respect to what ESPRESSO will provide. The Y band shows some strong telluric residuals that hinder the detection of the exoplanet water feature. However, the same band is accessible at higher resolving power (e.g. CARMENES), and weighted CCF methods (adapted from \citealt{Baranne1996, Pepe2002}) may enhance the signal.\\
It may be surprising that a signal is recovered in the J and H bands, considering the non-negligible absorption by molecular oxygen, carbon dioxide and methane there. This is due to the different structure of the spectrum of these species compared to water. Water has a dense forest of lines, while oxygen, carbon dioxide and methane have a more symmetrical structure which results in well separated lines (see Fig. \ref{Fig:zoom_oxygen_band}). At the resolution of GIANO, it is thus possible to use the information in-between the lines of these molecules to detect water, even in the absence of any weighting or masking of the telluric lines.

\subsection{Photon noise}
\label{appendix:photon}
To estimate the impact of photon noise on the retrieval of the contrast of the CCFs, we simulate observations of HD189733 with the ETCs of ESPRESSO\footnote{\url{https://www.eso.org/observing/etc/bin/gen/form?INS.NAME=ESPRESSO+INS.MODE=spectro}} and of GIANO\footnote{\url{http://tngweb.tng.iac.es/giano-b/etc/}}. We assumed:
\begin{itemize}
\item A K2V model for the host star, with magnitudes $\mathrm{V = 7.648}$ and $\mathrm{K = 5.541}$;
\item Seeing $1~\mathrm{arcsec}$, airmass 1.2, exposure time $300~s$;
\item Single HR 2x1 slow readout mode for ESPRESSO.
\end{itemize}
The ETCs provide the signal-to-noise ratio (SNR) on a stellar  spectrum in each resolution element, $\mathrm{SNR}_{\mathrm{res.\, elem.}}$. We then multiply this value for the square root of the telluric absorption template from \citep{Hinkle2003}, interpolating on a common grid. By doing so, we account for the increased photon noise within the core of the telluric lines. The ETC of GIANO partially accounts for telluric absorption, thus our estimate of the SNR is pessimistic. Finally, the signal-to-noise ratio $SNR(\lambda_\mathrm{obs})$ on each point of the simulated observations is obtained by
\begin{equation}
\begin{split}
\mathrm{SNR}(\lambda_\mathrm{obs}) = & \mathrm{SNR}_{\mathrm{res.\, elem.}}\sqrt{T_1(\lambda_\mathrm{obs})}\cdot\\
&\cdot\sqrt{\dfrac{\Delta \lambda_{\mathrm{obs}}}{\Delta \lambda_{\mathrm{res.\, elem.}}}}\sqrt{\dfrac{6\,480\cdot N_{\mathrm{transit}}}{300}}\,
\end{split}
\end{equation}
where $\Delta \lambda_{\mathrm{obs}}$ is the wavelength bin of the simulated observations (see Table \ref{tab:CCF_parameters}), $\Delta \lambda_{\mathrm{res.\, elem.}}$ is the size of the resolution element in wavelength (computed by the ESPRESSO ETC; $2.7~\mathrm{km~s^{-1}}$ for GIANO) and $N_{\mathrm{transit}}$ is the number of transits observed. We assumed that the transit duration is $6\,480~\mathrm{s}$ and neglected overheads. We also assumed that the baseline is observed at a much larger SNR than the transit. For HARPS, the SNR is lowered by $\sqrt{6.25}$ taking into account the different efficiency and size of the telescope.\\
These calculations take into account the spectral features of the host star and the decrease of flux in the cores of the individual telluric lines. Not surprisingly, the signal disappears within the \gianoone{}, \gianotwo{} and \gianothree{} bands. However, in the Y, J and H bands the signal is confidently recovered (see Fig. 
\ref{Fig:noisy_CCF}).\\
Table \ref{tab:errors_on_CCF} summarizes the results of our simulations. For each simulated cross-correlation function we compute the dispersion in its wings, which gives the error on the single resolution element in the velocity space. The results of our simulations are compatible with the precision obtained by \cite{Allart2017} on observations of HD189733b.\bigskip\\
Overall, provided that the planet is observed at sufficiently high systemic velocity and BERV, our analysis can be carried out with current ground-based instrumentation. With forthcoming instrumentation mounted at larger telescopes and with a broad spectral coverage, in particular at the ELTs, there is good hope that the photon and systematic noise will abated at the level desired for this kind of analysis.

\begin{table*}
\caption{We report the value of the dispersion on a single resolution element in the wings of the CCF of all considered bands for observations of a single transit. Note that the GIANO resolution element is larger, which is why the precision is higher. However, this results in a more poorly sampled CCF (see Fig. \ref{Fig:CCF_examples}).\label{tab:errors_on_CCF}}
\centering
\begin{tabular}{lccccc}
\hline 
Band name & Instrument simulated & Precision (1 transit) & Notes\\ 
\hline 
\\
\harps{} & HARPS(-N) & 51 ppm & \cite{Allart2017} reported 34 ppm with 3 transits,\\
&&& equivalent to 58 ppm with one transit,\\
&&& with a mask similar to that adopted in this paper\\
\rule{0mm}{0.4cm}
\espresso{} & ESPRESSO & 20 ppm & \\ 
\rule{0mm}{0.4cm}
\gianoone{} & GIANO & - & Accuracy limited by telluric contamination \\ 
\rule{0mm}{0.4cm}
\gianotwo{} & GIANO & - &  Accuracy limited by telluric contamination \\ 
\rule{0mm}{0.4cm}
\gianothree{} & GIANO & - &  Accuracy limited by telluric contamination \\  
\rule{0mm}{0.4cm}
Y band  & GIANO & 33 ppm & \\ 
\rule{0mm}{0.4cm}
J band & GIANO & 31 ppm &  \\ 
\rule{0mm}{0.4cm}
H band & GIANO & 27 ppm & \\  
\\
\hline 
\end{tabular}
\end{table*}

\begin{figure*}
\resizebox{\hsize}{!}{\includegraphics{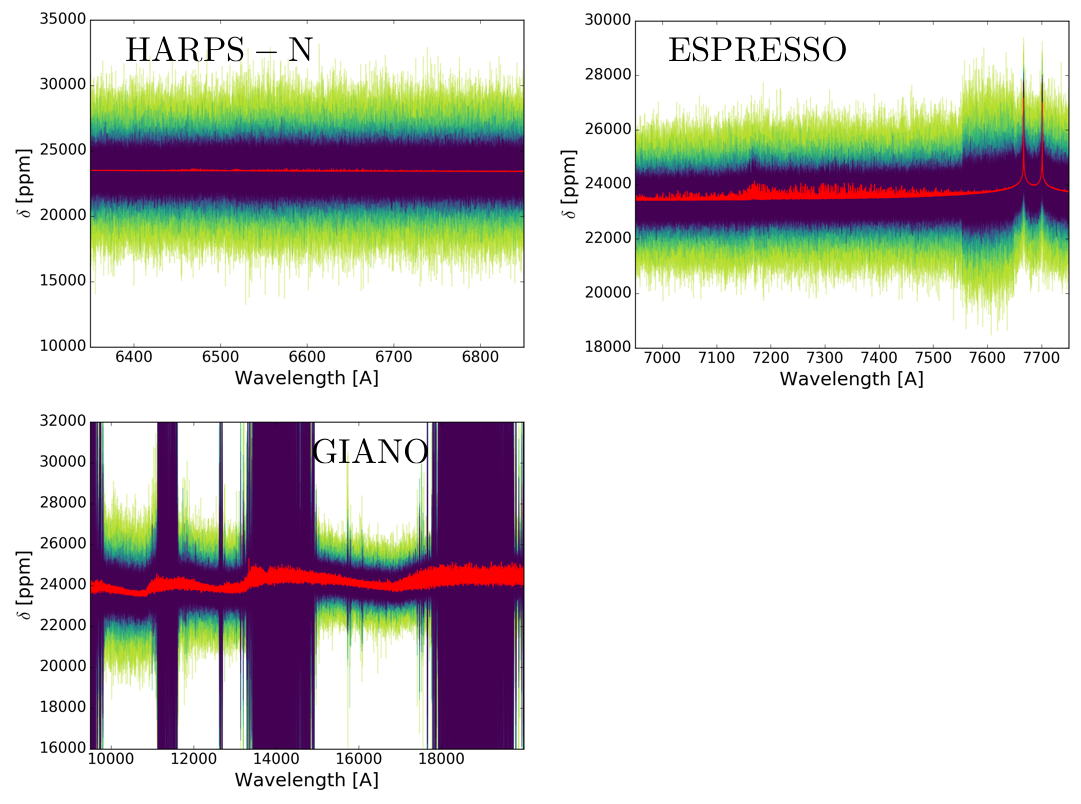}}
\caption{Simulations of observations of an aerosol-free HD189733b with HARPS-N (upper-left panel), ESPRESSO (upper-right panel) and GIANO (lower-left panel). The number of observed transits ranges from 1 (light green) to 9 (purple). A red solid line indincates a noise-free transmission spectrum. While the signal is buried in the photon noise, the combination of 800 water lines results in a detection in the \harps{}, \espresso{}, Y, J and H bands. In the \gianoone{}, \gianotwo{} and \gianothree{} bands, water absorbs all the incident light making detection impossible from the ground. \label{Fig:noisy_models}}
\end{figure*}

\begin{figure*}
\resizebox{\hsize}{!}{\includegraphics{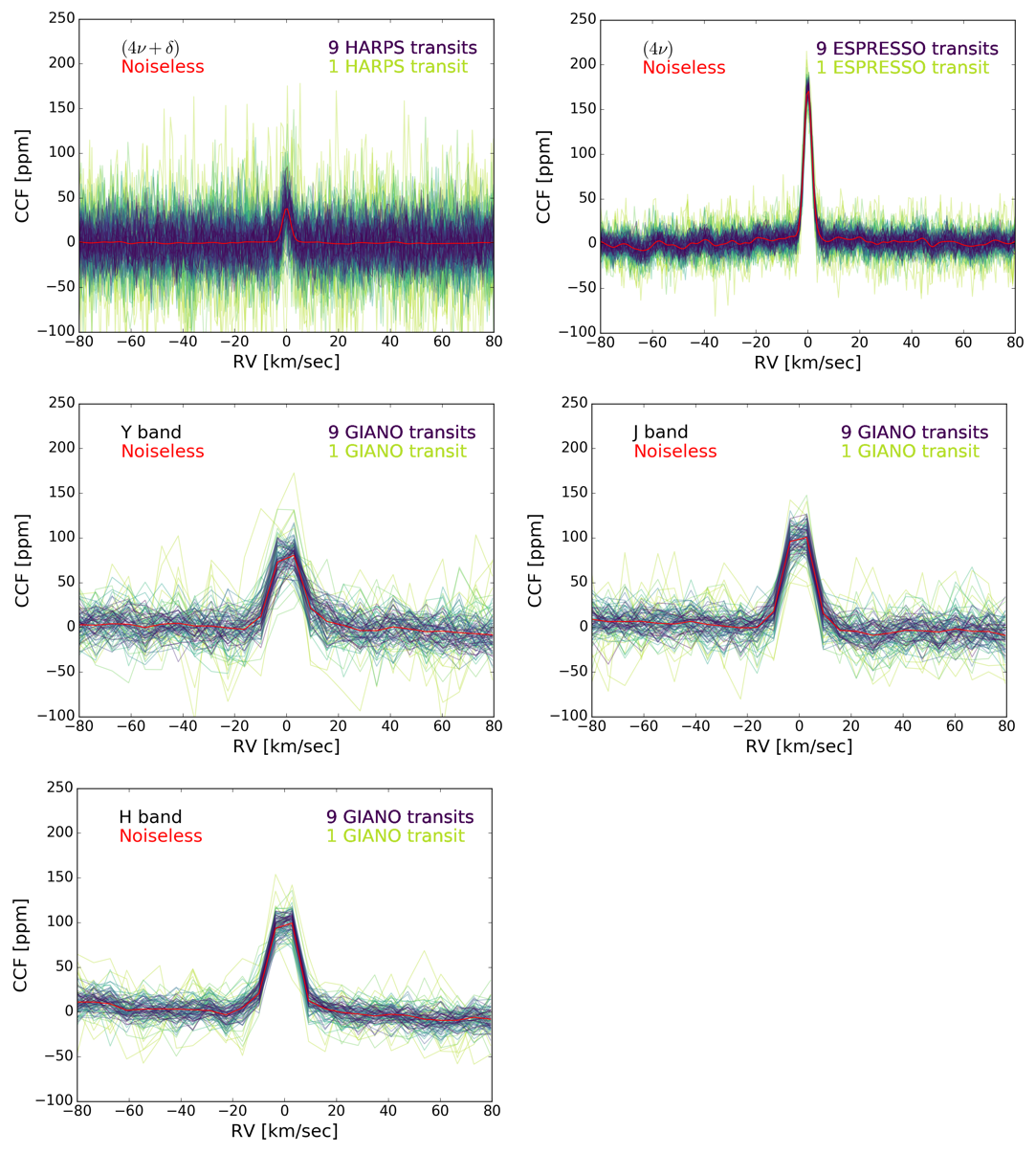}}
\caption{CCFs obtained measured from the models shown in Fig. \ref{Fig:noisy_models}. The dispersion values presented in Table \ref{tab:errors_on_CCF} are obtained in the wings of these CCFs. We do not show the \gianoone{}, \gianotwo{} and \gianothree{} bands where we tested that the CCF of water is not observable from the ground.\label{Fig:noisy_CCF}}
\end{figure*}

\begin{acknowledgements}
We thank the entire exoplanet atmospheres group of the Geneva Observatory and the Exoplanet and Stellar Populations Group of the University of Padova for insightful discussions. We also thank the anonymous referee, whose constructive critiques sensibly improved the quality of this paper. This work has been carried out in the frame of the National Centre for Competence in Research ‘PlanetS’ supported by the Swiss National Science Foundation (SNSF). L.P., R.A, D.E., C.L. and F.P. acknowledge the financial support of the SNSF. This project has received funding from the European Research Council (ERC) under the European Union’s Horizon 2020 research and innovation programme (grant agreement No 724427).
\end{acknowledgements}

\bibliographystyle{aa}
\bibliography{\centralbibliography}

\end{document}